# Influence of Molecular Organization on the Electrical Characteristics of π-conjugated Self-assembled Monolayers

Xavier Lefèvre,[1] Fabrice Moggia,[1] Olivier Segut,[1] Yu-Pu Lin,[2] Younal Ksari,[2] Grégory Delafosse,[2,3] Kacem Smaali,[4] David Guérin,[4] Vincent Derycke,[1] Dominique Vuillaume,[4] Stéphane Lenfant,[4*] Lionel Patrone,[2,3*] Bruno Jousselme[1*]

[1] CEA Saclay, IRAMIS, NIMBE/UMR 3685, Laboratory of Innovation in Surface Chemistry and Nanosciences (LICSEN), Gif-sur-Yvette Cedex F-91191, France.

[2] Institut Matériaux Microélectronique Nanosciences de Provence, Aix-Marseille Université, CNRS, Université de Toulon, IM2NP UMR 7334, Domaine Universitaire de St Jérôme, Service 151, 13397, Marseille Cedex 20, France

[3] Institut Supérieur de l'Electronique et du Numérique (ISEN-Toulon), CNRS, IM2NP UMR 7334, Maison des Technologies, Place Georges Pompidou, F-83000, Toulon, France

[4] Institute for Electronic Microelectronics and Nanotechnology (IEMN), CNRS, Univ. Lille, Av. Poincaré, F-59652 cedex, Villeneuve d'Ascq, France.

**Corresponding Authors**

* lionel.patrone@im2np.fr, stephane.lenfant@iemn.univ-lille1.fr and bruno.jousselme@cea.fr.

ABSTRACT

Two new thiol compounds with σ-π-σ structure were synthesized and self-assembled on gold substrates. The morphology and the structural characterization of SAMs assessed by infrared spectroscopy, contact angle, XPS, electrochemistry and scanning tunneling microscopy (STM) show the formation of monolayers. SAMs with a terthiophene (**3TSH**) core as conjugated system are much better organized compared to those with a naphthalene carbodiimide (**NaphSH**) core as demonstrated by the cyclic voltammetry and STM studies. The surface concentration of **3TSH** and **NaphSH** is respectively three and six times lower than ordered SAMs of pure alkyl chains. A large number of I/V characteristics have been studied either by STS measurements on gold substrates or by C-AFM on gold nanodots. Transition Voltage Spectroscopy (TVS) was used to clearly identify the transport in these partially organized monolayers. The chemical nature of the conjugated system, donor for **3TSH** and acceptor for **NaphSH**, involves an opposite rectification associated to the asymmetrical coupling of the molecular orbitals and the electrodes. The conductance histograms show that the **3TSH** junctions are less dispersed than those of **NaphSH** junctions. This is explained by a better control of the molecular organization in the molecular junctions.

**Introduction**

Organic thin films based on Self-Assembled Monolayers (SAMs) are particularly studied and envisaged for different applications such as bio compatibility, sensors, control of wetting and adhesion, catalysis, nanofabrication, nanoelectronics. In this last field, the use of SAMs made by the assembly of thiol compounds on gold is a key concept in a bottom-up approach to build active nanoelectronic devices that can switch, rectify, store and retrieve information.[1-4] The main advantages of these SAMs are that the monolayers are directly connected to a metal electrode, are easy to fabricate, handle and can be addressed to measure electrical properties. However metal-evaporation based techniques tend to damage molecular layers,[5] thus several methods have been developed to contact small assemblies of organized molecules without damaging them.[6] The main platforms and techniques investigated are the pioneering Hg-drop technique and the similar indium-gallium technique, the scanning probe microscope-based techniques as STM and conducting AFM (C-AFM), the soft-contact techniques such as nano-transfer printing or crossed wires and various solid-state device-based methods. Recently, the use of C-AFM was combined to sub-10 nm gold arrays to control and precisely measure the electronic transport properties through hundreds of molecules self-assembled on these nanodots.[7-8] This powerful method allowing to efficiently perform statistics on thousands of electrical measurements recording a single C-AFM picture was first applied to well-organized and structured linear alkyls chains.[8] However to improve the functionality of nano-devices, it is foreseen to develop more complex systems that introduce π-conjugated system as previously shown by the fabrication of molecular diodes,[9-13] switches and memories devices.[14-18] However, steric hindrance of the conjugated backbone modifies the interactions between molecules within the SAMs and is expected to reduce the degree of order within the molecular layer. To study and understand this behavior two thiol-terminated compounds with bulky conjugated systems named **3TSH** and **NaphSH** (Scheme 1) have been synthesized (see SI) and self-assembled, first on gold substrates to study the morphology and the structural characterization of SAMs by infrared spectroscopy, contact angle, XPS, electrochemistry and scanning tunneling microscopy (STM), then on gold nanodots to assess the electrical properties by C-AFM. The I/V characteristics obtained by C-AFM were compared with Scanning Tunneling Spectroscopy (STS) measurements. The STS I-V curves were then analyzed by Transition Voltage Spectroscopy (TVS).

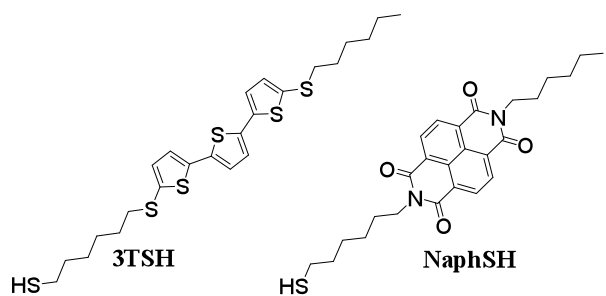

**Scheme 1.** Chemical structures of **3TSH** and **NaphSH**.

**Experimental section**

*Preparation of organic SAM–Au*. To perform XPS, IR spectroscopy and cyclic voltammetry, the gold substrates were prepared by the resistive evaporation of chromium (7 nm) followed by gold (200 nm) on silicon wafers at about $10^{-7}$ torr in a cryogenically pumped deposition chamber to obtain low roughness surfaces. For STM experiments, Au(111) surfaces on mica were purchased from Scientec. The gold nanodots (see SI for the fabrication) were first deprotected from the initial $SiO_2$ coating layer using HF then functionalized immediately with the thiol derivatives. All glassware used for the formation of monolayers was treated with piranha solution and rinsed with a large amount of high purity water (Millipore 18 MΩ.cm). *Caution: piranha solution reacts violently with most organic materials and should be handled with extreme care*. The freshly evaporated gold substrates, annealed or deprotected samples were immersed immediately after into a filtered $10^{-3}$ M solution of thiol in methylene chloride/ethanol mixture (ratio 1/4 for **NaphSH** and 0.5/9.5 for **3TSH**). Dodecanethiol solution ($10^{-3}$ M) dissolved in pure ethanol was used to make reference substrates. After 24 hours in the dark and at room temperature, organic SAMs-Au were rinsed carefully with methylene chloride then ethanol to remove the possible physisorbed overlayers and dried in a nitrogen stream. The various analyses were performed immediately after to avoid any deterioration and contamination of the surfaces. Prior to STM analyses, dodecanethiol SAMs on Au(111)/mica were annealed in ethanolic dodecanethiol solution at 60 °C for 2 h, followed by annealing in air at ~100 °C for 2 h.

*Apparatus for Characterizations:* XPS data were collected by an Axis Ultra DLD (Kratos Analytical) using a monochromatic Al Kα X-ray source (1486.6 eV). Infrared spectroscopy (IR) was carried out with a Bruker Vertex 70 spectrometer (MCT detector) equipped with a Pike Miracle plate for ATR. Electrochemical measurements were carried out with an EG&G potentiostat, model 273A. The values of contact angles with water were measured on Dataphysics – OCA15 EC apparatus, equipped with a CCD camera. Ellipsometry measurements were performed with a Sentech SE400 ellipsometer with 632.8 nm He-Ne laser at an incidence angle of 70° taking n=0.279 and k=3.37 for Au substrate (measured values) and n=1.45 and k=0 for organic layers. Results are averaged values measured from at least three different points on a substrate. STM experiments were carried out at ambient conditions using a *Multimode* system equipped with a *Picoamp* and a *Nanoscope IIIa* controller (Veeco/Bruker). The voltage was applied to the substrate with respect to the grounded Pt/Ir cut tips. The images were recorded in constant current mode, with a typical tunneling resistance $R_{th} \geq 100$ GΩ. For I(V) measurements, the same tip was used. STM results were analyzed and displayed using *WSxM* software.[19]

Ultraviolet Photoelectron Spectroscopy (UPS) was performed with a Omicron equipment (EA125 electron analyzer) and a UV source using the HeI line (21 eV) at normal incidence.[20] Inverse Photo-Emission Spectroscopy (IPES) spectra, giving access to the unoccupied electronic states, were recorded at normal incidence in the isochromat mode, detecting photons of around 9.7 eV. The absorbed current is kept at the order of 0.6 μA to prevent any degradation of the molecules. Details of the IPES working principle and experiment setups were described previously.[21] Conducting atomic force microscopy (C-AFM) was performed under a flux of $N_2$ gas (Dimension 3100, Veeco), using a PtIr coated tip. More detailed methods are provided in the SI.

**Results and discussion**

The combination of alkyl-chains and a π-conjugated system is a key design for the development of various molecular components. Thus, two new thiol derivatives (Scheme S1) with either a terthiophene or a naphthalene tetracarboxydiimide core enclosed between two linear hexyl chains have been synthesized. These aliphatic-aromatic-aliphatic structures (σ-π-σ) bear at one end of one alkyl chain a thiol function for the anchoring on gold substrate. The choice of alkyl chains with six carbons was driven to limit the coupling of the conjugated system with the gold electrode while enabling the transfer of electron through the π-system by tunneling. The two conjugated systems were chosen firstly for their different steric hindrance that should impact on their self-assembly, then for their opposite electronic nature with a donor or acceptor system for **3TSH** and **NaphSH**, respectively. In the following, this difference will help to discriminate between transport through the organized lattice or through defects in the SAMs.

Theoretical calculations based on density functional methods have been performed for the thiols-terminated compounds in the gas phase using the Gaussian 98 software.[22] Becke's three-parameter gradient-corrected functional (Becke3lyp) with polarized 6-31G* basis for all atoms was used to optimize the geometry and to compute the electronic structure of the compounds. The minimized conformations are presented in figure S1 (in SI). For both compounds the alkyls chains are out of plane of the conjugated system due to heteroatoms. Indeed, in the **3TSH** compound the two sulfurs in position 5 on terthiophene and in **NaphSH** the nitrogens are involved respectively a dihedral angles around 85° ($S_{thio}$-$C_5$-S-$CH_2$) and 90° ($C_{Carb}$-N-$CH_2$-$CH_2$). The length of **3TSH** between the carbon of the methylene and the sulfur of the thiols groups is 27.7 Å, and in the case of **NaphSH** the length is shorter and is 23.2 Å. Concerning the conjugated systems, the terthiophene is almost twice as long and its width is smaller than the naphthalene tetracarboxydiimide core.

XPS analysis of SAMs of **3TSH** and **NaphSH** on Au (figure S2 in SI) show on the survey spectra the presence of C and S for the terthiophene derivative and C, N, O and S for the naphthalene compound in agreement with the chemical structures. The ratio of the different elements calculated by the deconvolution of the XPS spectra regions is also in agreement with the elemental composition of the compounds (Table S1 in SI). XPS spectra in the $S_{2p}$ region (insets in figure S2) displays for **NaphSH** two peaks with a bonding energy at 162.0 eV and 163.2 eV corresponding to $2p_{3/2}$ and $2p_{1/2}$ with a ratio of 2:1 as theoretically determined by spin-orbit splitting.[23-24] Those two peaks are assigned to one specie, i.e., the thiolate (bond sulfur) coming from the oxidant addition of the thiol on gold to form a metal-sulfur bond. Similarly for **3TSH,** peaks at 162.0 eV and 163.2 eV show the thiolate link with the gold and the other two peaks at 163.8 and 165.0 eV correspond to the other five sulphurs of the compound. Contact angles measured with water droplet on the self-assembled monolayers of **3TSH** and **NaphSH** are between 105 and 108° and between 100 to 105°, respectively. As a comparison, SAMs of dodecanethiol, made on pure methanol using the same substrates and conditions, exhibit contact angles of 110°, in agreement with the literature.[25] Thus, contact angles with SAMs of **3TSH** are slightly better than the **NaphSH** and very close to the one of dodecanethiol. Thickness of **3TSH** and **NaPhSH** SAMs on Au was measured by ellipsometry as ~25.7 Å and ~20.8 Å respectively, in good agreement with the molecule length and consistent with a single layer of *all-trans* molecules on Au surface with respective tilts of 22° and 26°. Since the molecules bear alkyl chains at both ends, it is not surprising that these values are in a similar range of those of 26-28° concerning alkylthiols on Au(111).[26] Infrared spectra of thiol derivatives were recorded in carbon tetrachloride solution or chemisorbed on gold with single reflection attenuated total reflectance (ATR) technique using gold substrate as a mirror (figure S3 in SI). For both compounds, the molecules organized as monolayers present different absorption spectra than those dissolved in a solution. This is shown for example by an increase in intensity of the symmetric (~ 2875 $cm^{-1}$) or anti-symmetric (~ 2955 $cm^{-1}$) stretching of the $CH_3$ end group compared to the $CH_2$ stretching of the aliphatic chains (Symmetric at ~ 2855 $cm^{-1}$ and anti-symmetric at ~ 2925 $cm^{-1}$). These results comparable to those obtained on similar σ-π-σ self assembled monolayers[27] with Reflection–Absorption infrared spectroscopy (RAIRS) techniques can be explained by the surface dipole selection rule on metal surfaces that enhances the intensity of the vibrational modes perpendicular to the surface and inhibit the electric dipole moment aligned parallel to the plane of incidence. Thus, the increase in intensity of the methyl group mode and the decrease of the methylene group imply that the two thiol compounds stand almost normal to the substrate surface. Similar variations in the conjugated part of **NaphSH** can be seen and particularly on the carbonyl groups of the imides functions which present in dilute medium higher intensity for asymmetrical vibration at 1660 cm$^{-1}$ than the symmetrical one at 1705 cm$^{-1}$.[28] In condensed medium, the intensities are reversed and the symmetrical vibrations are more intense. These intensity variations are less pronounced in the terthiophene core (1495, 1431 cm$^{-1}$) but new vibrational modes appear at 1266 or 1100 cm$^{-1}$ for **3TSH** as well as at 1340 cm$^{-1}$ for **NaphSH** due to the breaking of the symmetry of the molecules chemisorbed on the gold substrates. Thus, it can be concluded that the two organic compounds stand almost normal to the surface and are not lying down.

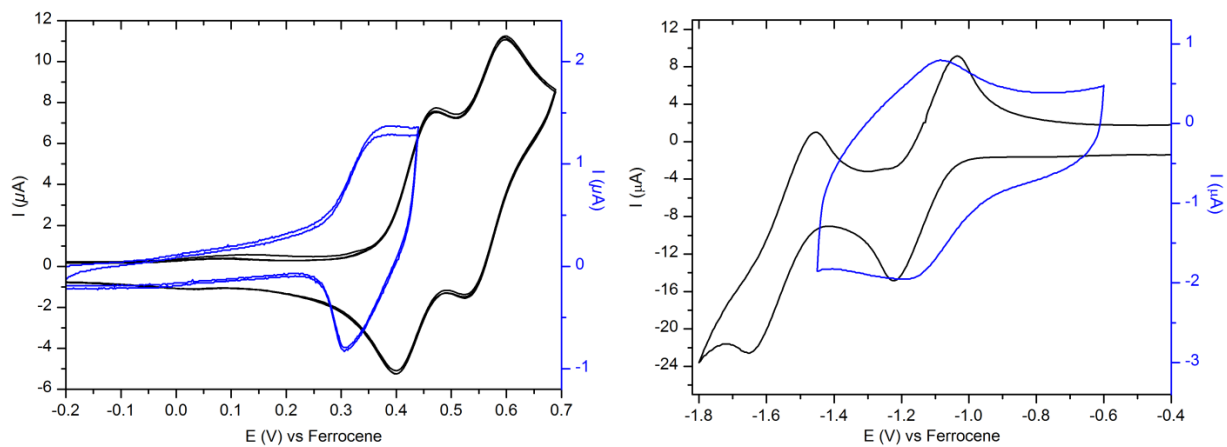

**Figure 1.** Cyclic voltammetry on a gold electrode of millimolar solutions (black curves) of references **3THex** (left) or **NaphHex** (right) in CH$_2$Cl$_2$-Bu$_4$NPF$_6$ 0.1 mol.L$^{-1}$ at 100 mV.s$^{-1}$ or of self-assembled monolayers on gold (Blue curves) of **3TSH** (left) or **NaphSH** (right) in CH$_2$Cl$_2$-Bu$_4$NPF$_6$ 0.1 mol.L$^{-1}$ at 100 mV.s$^{-1}$.

Cyclic voltammetry studies of 5,5''-dihexyl-2,2':5',2''-terthiophene (**3THex**) and N,N'-dihexylnaphthalene-1,8-dicarboxyanhydride-4,5-dicarboximide (**NaphHex**) reference compounds (see SI for the chemical structures) presenting no thiol function show respectively two reversible oxidation and reduction processes on gold working electrode (black curves in figure 1). The *alpha* positions locked with the two hexylsulfanyl substituents give to **3THex** the abilities to remove and re-inject two electrons in the rich electron conjugated system without any radical polymerization. In the other case, inductive effect of the four carbonyl electron withdrawing units distributed symmetrically around the π-conjugated naphthalene core allows to inject reversibly two electrons in the conjugated part. For **3TSH** or **NaphSH** chemisorbed on gold, respectively, only one oxidation and reduction reversible wave can be analyzed in a range from -1.45 to 0.45 V vs ferrocene (blue curves in figure 1). Outside this window, these reversible redox waves completely disappear. Indeed, at lower or higher potential than -1.45V or 0.45V vs ferrocene respectively, the desorption of the SAMs proceeds as previously seen for alkane thiol SAMs.[29-30] The cyclic voltammograms (CVs) of **3TSH** or **NaphSH** self-assembled on gold show smaller Epa–Epc compared to those of the **3THex** and **NaphHex** reference compounds in solution. This is consistent with the fact that thiol compounds are surface bound (chemisorbed) redox species. The intensities of anodic or cathodic peaks for the SAMs are also directly proportional to the scan rate (figures S4 and S5 in SI). The surface coverage of **3TSH** and **NaphSH** in the SAMs was estimated from the charge measured respectively for its oxidation during the anodic sweep or the reduction during the cathodic sweep in an electrolyte free of thiols. Values of $2.10 \times 10^{-10}$ mol.cm$^{-2}$ ($1.3 \times 10^{+14}$ molecules.cm$^{-2}$ or about 80 Å$^2$ per molecule) for **3TSH** SAM-Au and $1.25 \times 10^{-10}$ mol.cm$^{-2}$ ($7.6 \times 10^{+13}$ molecules.cm$^{-2}$ or about 130 Å$^2$ per molecule) for **NaphSH** SAM-Au were obtained. This represents almost three to six times lower surface concentrations than ordered pure alkyl chain SAMs with $7.6 \times 10^{-10}$ mol.cm$^{-2}$ (ca. 22 Å$^2$ per molecule).[31] It can be explained by steric effects of the conjugated part of the thiol compounds that decrease density of molecules and involve less compacity and density in the self-assembled monolayers. The **NaphSH** chemisorbed on gold shows similar redox potential for the first reversible peak than those of the **NaphSHex** in methylene chloride solution (table S2 in SI) showing no interactions between the naphthalene tetracarboxydiimide in the SAMs. In the other case, the oxidation peak of the terthiophene in **3TSH** SAM-Au shifts by - 95 mV compared to those in **3TSHex** in solution, this cathodic shift is explained by π-π interactions of the conjugated system that ease their oxidation. Moreover, **NaphSH** SAM exhibits considerably more double layer capacitance than the **3TSH** SAM in the electrochemical response. Higher capacitive currents come from the counter-ion insertion into SAMs during electrochemical polarisation .[32] Thus, the observation of higher capacitance in the **NaphSH** SAM shows that this monolayer is less compact than the **3TSH** SAM. The highest occupied molecular orbital (HOMO) and the lowest unoccupied molecular orbital (LUMO) energy levels of the thiol derivatives can be determined using the described empirical relationships established : $E_{HOMO} = -(E_{ox} + 4.4)$ eV and $E_{LUMO} = -(E_{red} + 4.4)$ eV with $E_{ox}$ and $E_{red}$ potential values determined in solution vs. SCE reference electrode.[33] Ferrocene, used as reference in CV experiments in figure 1, has a potential of 0.48 V vs. SCE in methylene chloride electrolyte.[34] Using the formula, the HOMO level of terthiophene derivative is determined at -5.31 eV using $E°ox = 0.91$ V *vs*. SCE ($E°ox = 0.43$ V *vs*. ferrocene, see table S2). The LUMO level of naphthalene tetracarbodiimide is estimated at - 3.75 eV using $E°_{red} = -0.65$V *vs*. SCE ($E°_{Red} = -1.13$V *vs*. ferrocene, table S2).

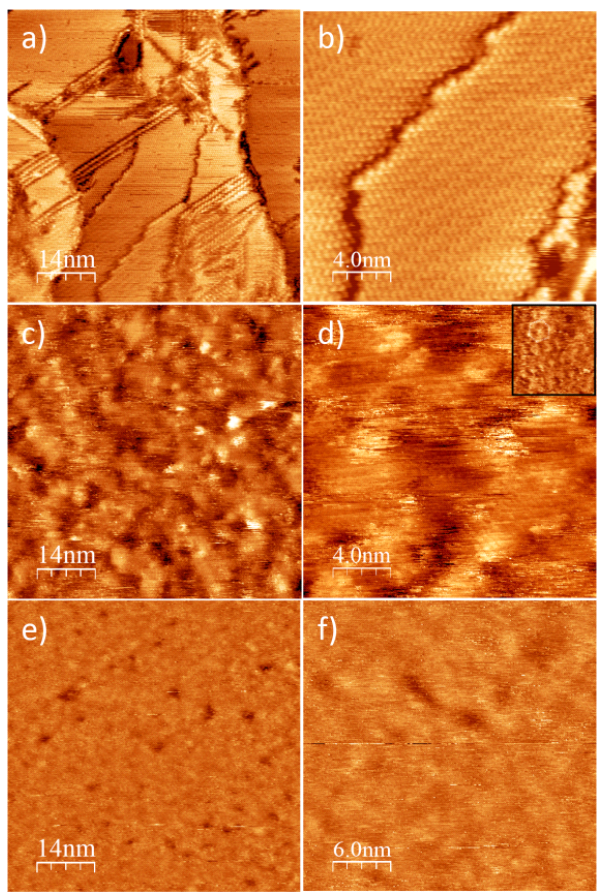

**Figure 2**. STM Images of: a) & b) **C$_{12}$SH** SAM-Au (-0.7 V, 2 pA), c) & d) **3TSH** SAM-Au (-0.9 V, 20 pA), and e) & f) **NaphSH** SAM-Au (-0.9 V, 2 pA & 20 pA respectively) on Au(111) evaporated on mica. Inset of d) is a 4.6×4.2 nm² zoom on **3TSH** SAM showing hexagonal molecular arrangement. Topographic Z scale is 1.5 nm from dark to bright for a), c), d), e) & f), and 0.5 nm for b).

STM images of dodecanethiol (**C$_{12}$SH**), **3TSH** and **NaphSH** SAMs on Au(111) on quartz are presented in figure 2. Pictures of **C$_{12}$SH** SAMs (figure 2a and 2b) are similar to the ones reported in the literature[35-38] with close-packed organized molecular domains separated by grain boundaries. A typical arrangement is *(√3 ×√3) R30°* hexagonal structure with intermolecular distance of ~5 Å. As for **C$_{12}$SH**, **3TSH** and **NaphSH** SAMs exhibit "etch pits" in the gold surface typical for thiolate grafting. Adsorption of a single monolayer was further confirmed by the depth values of holes induced by a pulse voltage applied between the tip and the surface: ~22 Å for **NaphSH** and ~30 Å for **3TSH**, i.e. almost similar to the respective molecular length. Moreover, concerning **3TSH** Au SAMs (see figure 2c and 2d), ordered small molecular areas of about 10×10 nm² can be seen all over STM images. However, these arrangements are also separated by obviously more disordered zones that prevent from obtaining a molecular resolution as clear as for **C$_{12}$SH** Au SAMs in figures 2a and 2b. Zoomed image presented in figure 2c inset shows that **3TSH** molecules are separated by ~5 Å within an hexagonal lattice, like for **C$_{12}$SH**. Such an organization has already been reported[39] for SAMs on gold of terphenyl molecules bearing a small alkyl chains. In the same way, long alkyl chain of **3TSH** gives the molecules the necessary flexibility for their assembly and their ordering via π-π interactions of the conjugated system as demonstrated by cyclic voltammetry. Nevertheless, in spite of their similar alkyl chains, whatever the STM imaging conditions we could not observe any molecular organization for **NaphSH** SAMs (figure 2e and 2f). In this case, the conjugated part of the molecule may be too wide for enabling a close-packed arrangement and/or possible C–H···O=C hydrogen bonds may compete with ordering of grafted molecules.[40]

To study the electrical transport through SAMs of **3TSH** and **NaphSH**, a statistical analysis on a large number of devices is required. Thus, the recently reported new approach that allows to measure the conductance of up to a million of junctions in a single C-AFM image was investigated.[7] Molecular junctions of **3TSH** and **NaphSH** were fabricated on a large array of sub-10 nm single crystal Au nanodot electrodes; each nanodot/molecules/tip junction contains less than one hundred molecules. An array of gold nanodot electrodes was fabricated by e-beam lithography and lift-off technique. Each nanodot is embedded in a highly-doped Si substrate (to insure a back ohmic contact), covered by a SAM of molecules of interest and contacted by the C-AFM tip (see SI, Figure S6a). The main properties of the nanodots relevant for the molecular conductance measurements are briefly recalled: The distance between each nanodot is set to 100 nm. The estimated nanodot diameter measured from scanning electron microscopy (SEM) and high resolution transmission electron microscope (HR TEM) is 8 nm at the interface with Si and 5 nm at the top surface.[7] The height is 7 nm.[7] Using the surface coverage determined above by electrochemistry on gold, each nanodot junctions contains either ~25 molecules of **3TSH** or ~ 15 molecules of **NaphSH** sandwiched between the top Au nanodot surface and the C-AFM tip. By sweeping a C-AFM tip at a given bias (applied on the substrate), current was measured only when the tip is on top of the molecular junction since conductance of native $SiO_2$ is below the detection limit of our apparatus.[8] The loading force of the C-AFM tip is kept very low (1 nN) to avoid any deformation of the SAM and the strain-induced modification of their electronic properties.[41] Figures S6b and S6c (see SI) show typical C-AFM images taken at -0.8 V and +0.8 V, respectively, for 2280 Au nanodot/**NaphSH**/C-AFM tip molecular junctions. Using a thresholding program,[8] the current histograms were constructed and shown in figure 3 for voltages -0.8V and +0.8 V. These histograms were well fitted by several log-normal distributions. In the framework of a non-resonant tunneling transport through the molecular junction, the current is exponentially dependent on the SAM thickness and on the interface energetics (i.e. position of the molecular orbitals relatively to the electrode Fermi energy), thus any normal distribution of these parameters leads to a log-normal distribution of the conductance as already observed in many molecular junctions.[5,12-13,18,42-47]

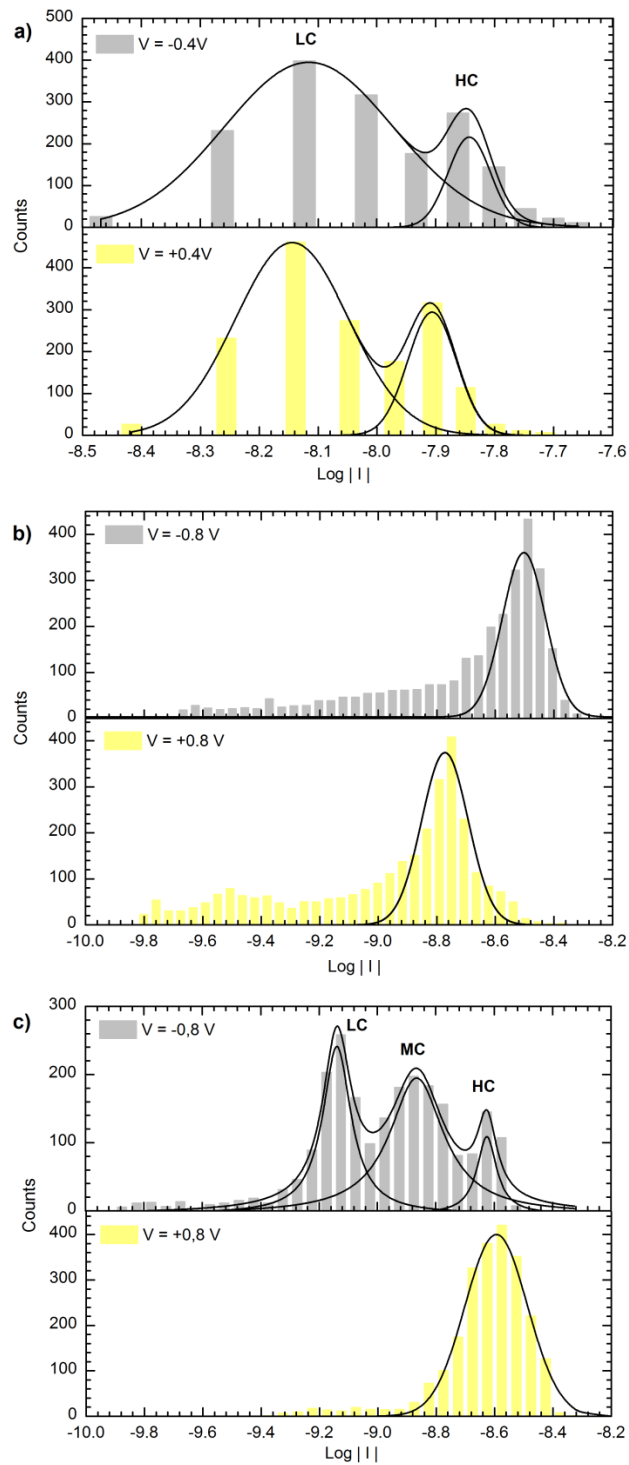

**Figure 3.** Histograms of the current for a) **C$_{12}$SH** (1639 junctions), b) **3TSH** (2993 junctions) and c) **NaphSH** (2280 junctions) molecular junctions on a nanodot electrode at a fixed bias of -

0.8 V (grey) or +0.8 V (yellow), except for $C_{12}SH$ measured at ±0.4V (Fig. 3a, reprinted (adapted) with permission from Ref. 6. Copyright 2012 American Chemical Society).

Figures 3-a, b and c show the current histograms recorded by the method described above for $C_{12}SH$, **3TSH** and **NaphSH** molecules, respectively. The parameters (mean current and standard deviation) of the fitted log-normal distributions are given in table S3 (see SI). For **NaphSH** junctions (figure 3c), at negative voltage, the histogram was decomposed into three peaks (labeled HC, MC and LC for "high conductance", "medium conductance" and "low conductance", respectively), while at positive voltage, only one peak was observed. The populations of the three peaks at -0.8V are ~42% (955 junctions), ~43% (983 junctions) and ~15% (342 junctions) for the LC, MC and HC peaks, respectively. The repartition was obtained by summing the junction counts belonging to each peak and normalizing by the total of 2280 junctions measured in that case. We note that the HC peak at -0.8 V and the peak at +0.8 V have almost the same mean value of $2.3$-$2.4 \times 10^{-9}$ A. Thus, we can infer that about 15% of the **NaphSH** junctions do not show any rectification, while 85% show a weak positive rectification, i.e. $R_+ = I_+/I_-$ is larger than 1 where $I_+$ is the mean current of the peak at +0.8V (see data in Table S3) and $I_-$ is the mean current of the peaks at -0.8V. Out of this 85%, 43% (the MC peak) show a weak rectification with $R_{+MC} = I_+/I_{-MC}$ ~1.6 and 42% (the LC peak) have a rectification ratio $R_{+LC} = I_+/I_{-LC}$ ~3.2. For the **3TSH** junctions, the shape of the histograms is quite different compared to **NaphSH**. At both negative and positive voltages, ~80% of the junctions belong to a main peak. The rest appears as a shoulder and a long tail at lower currents. The histogram at -0.8V is shifted toward higher currents than at +0.8V, corresponding to a negative rectification, with a rectification ratio $R_- = I_-/I_+$ ~1.9 (See data in SI, Table S3). This is consistent with our present

observation of a better organization in the SAMs for the **3TSH** compared to **NaphSH**. For the reference **C$_{12}$SH** the histograms are symmetrical (no rectification). The bi-modal distribution on figure 3a has been ascribed to the existence of different molecular organization phases for such alkylthiol monolayers.[6]

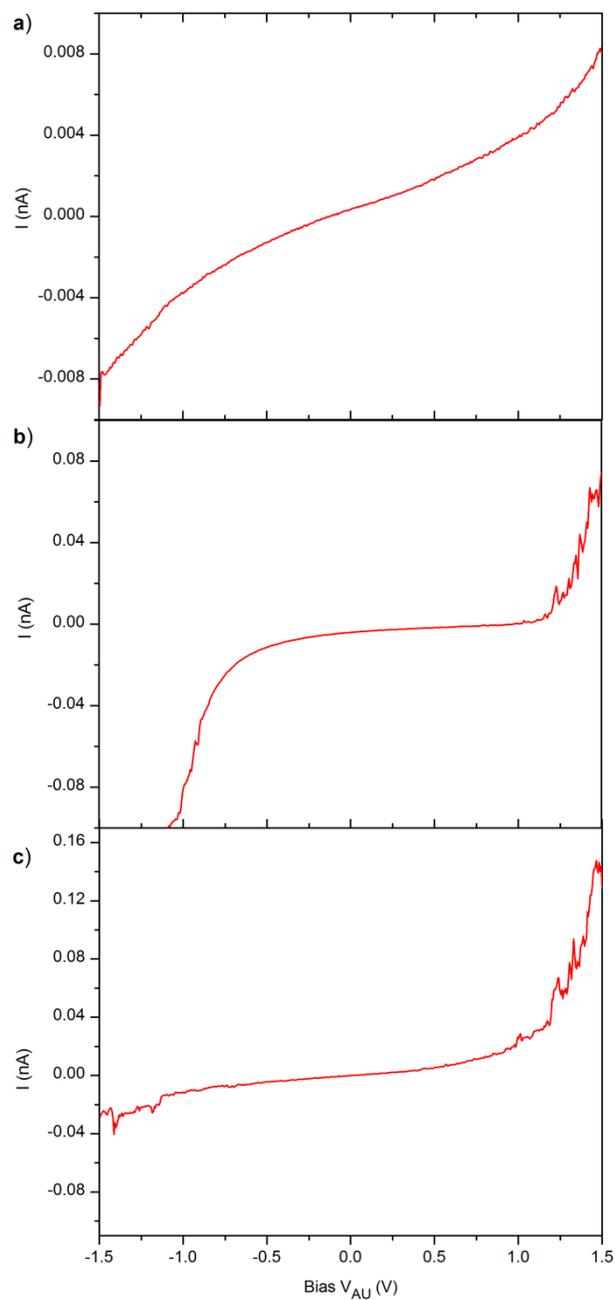

**Figure 4**. I(V) measurements performed with the same STM tip on a) **C$_{12}$SH**, b) **3TSH** and c) **NaphSH**. Each curve is an average of five single measurements cycled from -1.5 V to +1.5 V. STM settings are a set-point current of 2 pA, and a bias of -0.7 V for a) and c) and +0.7 V for b).

Local electrical measurements of the SAMs on flat Au(111) deposited on mica have been also performed using the STM tip. Results are presented in figure 4. About 70-80% of measured I(V) curves exhibit a characteristic behavior related with intrinsic electrical properties of the molecules. For **C$_{12}$SH** SAM, I(V) curve is rather symmetrical with a low current level (figure 4a). This rather symmetrical behavior is in agreement with the C-AFM measurements reported in Fig. 3a. As done from the C-AFM results, rectification ratios $R_+ = I_+/I_-$ or $R_- = I_-/I_+$ have been investigated from the whole STM I(V) curves at ±0.8V. Histograms of rectification ratios of **3TSH** and **NaphSH** are presented in figure S7 (in SI). The **3TSH** exhibits a sharp single rectification ratio at negative bias $R_- = I_-/I_+$ with a mean value at ~7.8. Although the rectification ratio $I_-/I_+$ is higher than that measured by C-AFM (~1.9), both measurements give a single distribution for **3TSH** thus indicating rather well controlled electrical properties for **3TSH** SAMs. Conversely, the rectification ratio $R_+ = I_+/I_-$ of **NaphSH** SAMs exhibit a wider distribution with a mean value 1.3, extending up to about 10 (see Fig. S7, in SI). These values are consistent with those determined (see above) from the current distribution measured by C-AFM at ±0.8 V (Fig. 3c). For a donor molecule, as **3TSH**, a rectifying behavior at negative bias (Figures 3b and 4b) is expected if the HOMO is asymmetrically coupled to the electrodes (see Figure S8, in SI), i.e. if the coupling to the Au electrode $\Gamma_{Au}$ is larger than to the tip $\Gamma_{tip}$. Here, the Au-thiol bond and the short C$_6$ chain determine $\Gamma_{Au}$. It is reasonable assuming the same for both C-AFM and STM experiments. The coupling parameter to the STM tip, $\Gamma_{STMtip}$, depends on the C$_6$ chain and the tunnel air gap between the tip and the molecules, thus it is likely that $\Gamma_{STMtip} < \Gamma_{Au}$ leading to the observed rectification. In the C-AFM measurements, the tip is gently on contact with the molecules, thus $\Gamma_{CAFMtip}$ increases (no tunnel air gap) compared to $\Gamma_{STMtip}$ reducing the rectification behavior as observed from C-AFM ($R_- \sim 1.9$) (Figure 3b) compared to STM ($R_- \sim 7.8$) (Figure 4b). Similarly, for an acceptor molecule as **NaphSH**, a rectification at positive bias is expected via the LUMO level asymmetry coupled to the electrodes (Figure S8, SI). However, due to the multimodal current distribution in C-AFM (Fig. 3c) and the large distribution of $R_+$ in STM, it is more difficult to apply the same discussion as above in this case. A more detailed discussion will be given below regarding the electronic structure of these **NaphSH** SAMs.

To assess the energy level position of the molecular orbital involved in the electron transport through the junction, we analyzed the STS I-V curves by Transition Voltage Spectroscopy (TVS). With this approach, the voltage $V_T$ at the minimum of the Fowler-Nordheim plot ($\ln(I/V^2)$ *vs* $1/V$) is proportional to the energy barrier[48-49] at the molecule/electrode interface, i.e. the energy gap between the Au Fermi energy and the molecular orbital involved in the charge transport (see figure S9 in SI). The histograms of transition voltage values extracted from ~100 (**C$_{12}$SH**), ~140 (**NaphSH**) and ~250 (**3TSH**) I-V curves are plotted in figure 5.

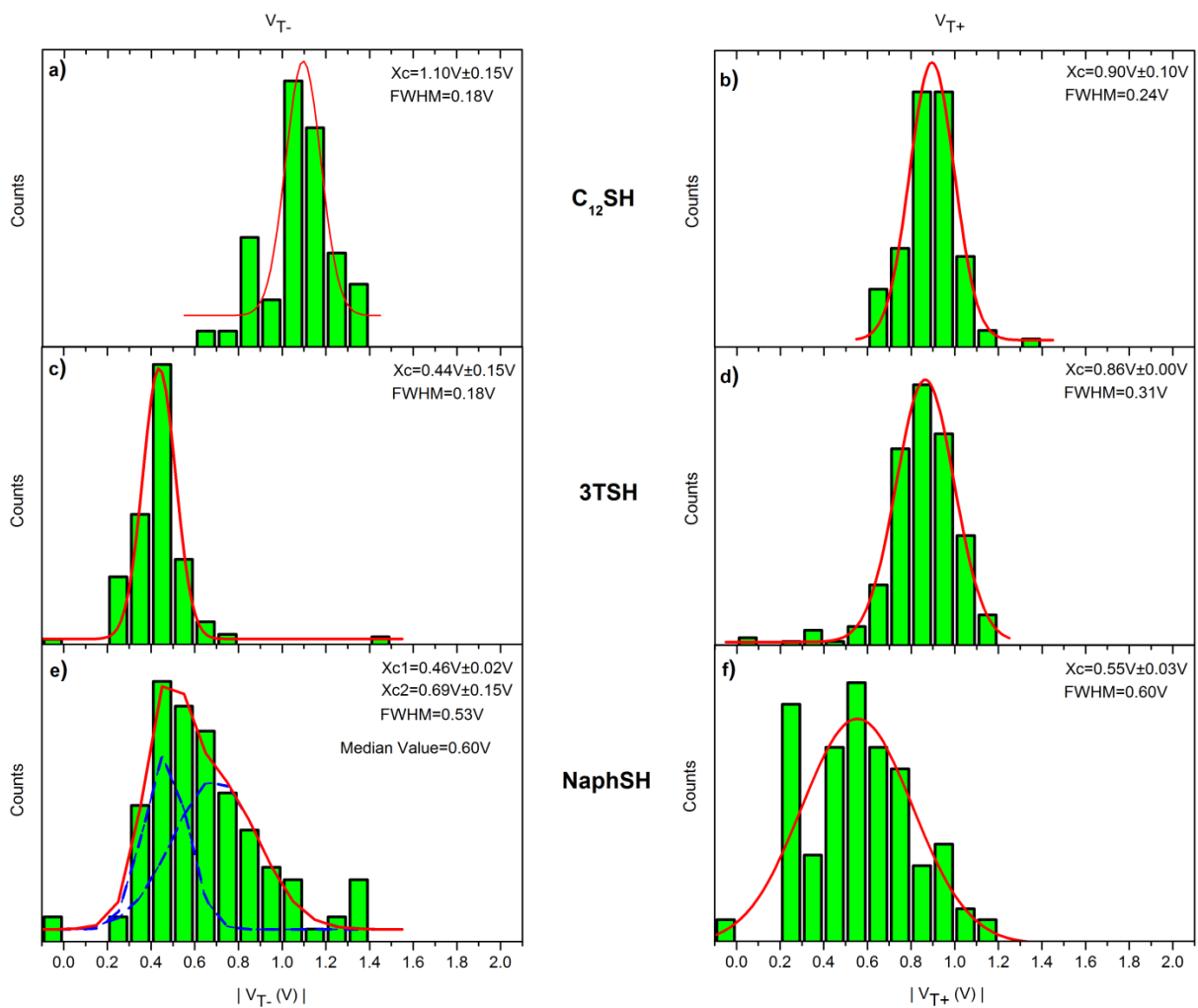

**Figure 5.** Transition voltage distribution (absolute values) at negative ($V_{T-}$) and positive ($V_{T+}$) bias for **C$_{12}$SH** (a) and b)), **3TSH** (c) and d)), and **NaphSH** (e) and f)). Distributions were fitted by Gaussian curves centered at Xc values. $V_{T-}$ distribution for **NaphSH** SAM is best fitted using two Gaussian curves. Full width at half maximum values (FWHM) are also reported.

For **C$_{12}$SH** and **3TSH,** transition voltage distributions are narrower and Gaussian fits enable to extract a mean voltage value. On the contrary, for **NaphSH,** distributions are broader, and two $V_{T-}$ can be deduced (Figure 5e). For each couple of $V_{T-}$ and $V_{T+}$ values assuming that the electronic transport occurs through a single molecular orbital, it is possible to calculate the position $\varepsilon_0$ of that level with respect to the Fermi level using a model proposed by Bâldea:[48]

$$|\varepsilon_0| = 2\frac{e|V_{T+}V_{T-}|}{\sqrt{V_{T+}^2 + 10|V_{T+}V_{T-}|/3 + V_{T-}^2}}$$

(Eq. 1)

The calculated values are given in table I. The $V_T$ values (about 1.0V) and the $\varepsilon_0$ of ~0.9 eV for **C$_{12}$SH** are consistent with literature data.[7,49-51] Albeit values for **3TSH** and **NaphSH** were not previously reported, the $V_T$ and $\varepsilon_0$ values found here are consistent with the expected lower values for π-conjugated molecules as also observed for various phenyl derivatives.[49] To check which orbital (HOMO or LUMO) is involved in the transport, we used photoemission spectroscopy. Indeed, UPS allows determining the HOMO position whereas the LUMO can be accessed using IPES (Fig. 6). The HOMO and LUMO positions were extracted measuring their onset regarding the Fermi level as proposed in the literature.[52]

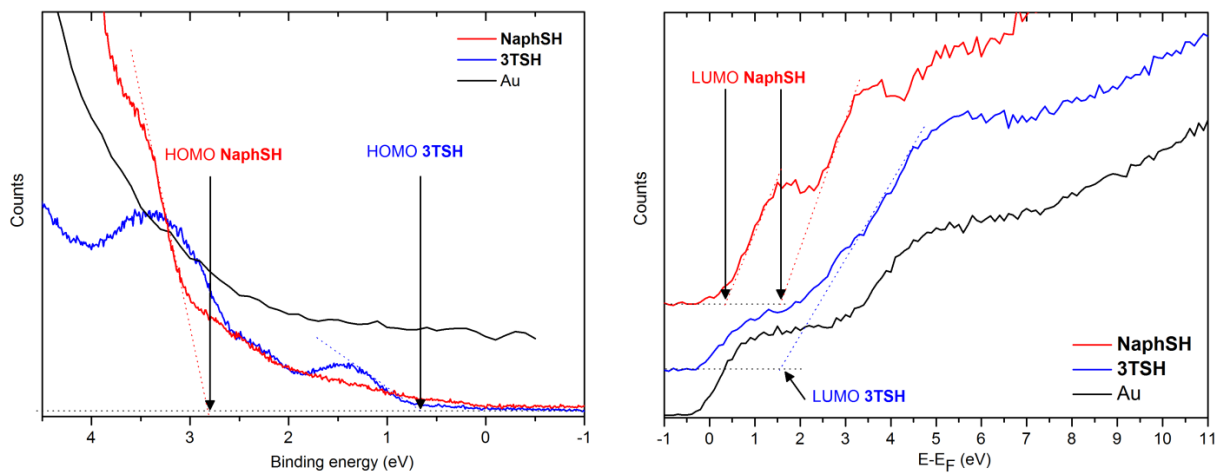

**Figure 6.** UPS (left) and IPES (right) spectra of **3TSH** and **NaphSH** SAMs on gold. For the sake of clarity IPES spectra were offset along the z axis. Estimated values of HOMO and LUMO

were determined by the intersection of the tangent to the band (dotted line) with the spectrum baseline as indicated by the arrows. To bring out LUMO position, IPES spectra with subtracted background signal are plotted in SI (see Figure S10).

The whole results obtained from STM I(V), CV and UPS/IPES are summarized in Table I. As expected the **3TSH** has the HOMO closer to the Fermi energy than the LUMO (donor molecule), while it is the reverse situation for **NaphSH** (acceptor molecule).

|  | $V_{T-}$ (V) | $V_{T+}$ (V) | $\varepsilon_0$ (eV) | $\gamma$ | HOMO (eV) (vs vac.) | LUMO (eV) (vs vac.) | HOMO (eV) from CV (vs. vac.) | LUMO (eV) from CV (vs. vac.) |
|---|---|---|---|---|---|---|---|---|
| **C$_{12}$SH** | -1.10 | 0.90 | 0.86 | -0.04 |  | ~1.00* |  |  |
| **3TSH** | -0.44 | 0.86 | -0.51 | -0.14 | -0.65 ± 0.05 (-5.50 ± 0.05) | 1.60 ± 0.10 (-3.30 ± 0.10) | -5.31 |  |
| **NaPhSH** | -0.46 | 0.55 | 0.43 | 0.04 | -2.80 ± 0.05 (-4.50 ± 0.05) | 0.40 ± 0.05 |  | -3.75 |
|  | -0.69 |  | 0.53 | -0.05 | (-7.70 ± 0.05) | 1.60 ± 0.05 (-3.30 ± 0.05) |  |  |

**Table I.** Transition voltage values, calculated position of the molecular energy level $\varepsilon_0$ (from Bâldea, Eq. 1), calculated $\gamma$ (Eq. 2), HOMO/ LUMO positions obtained from UPS/IPES with respect to the Fermi level, and into brackets with respect to the vacuum level (taking a work function of 4.9 eV for Au), and HOMO/LUMO determined from CV measurements. *For **C$_{12}$SH**, LUMO position corresponds to the onset of the band tail reported in IPES spectra of **C$_{18}$SH** in the literature.[53]

For the **C$_{12}$SH**, the $\varepsilon_0$ value (~0.9 eV) is consistent with the onset of the LUMO tail as measured in similar Au/alkanethiol systems by IPES,[53] and consistent with previous TVS measurements in similar Au/C$_{12}$SH/Au molecular junctions.[51] For **3TSH**, there is a good agreement between $\varepsilon_0$ (- 0.51 eV) values deduced from TVS analysis and the position of the HOMO measured by UPS (-0.65 eV), this last value being also in agreement with the CV measurements (ionization potentials: $IP_{UPS}$=-5.55eV, $IP_{CV}$=-5.3 eV). For **NaphSH**, since we have two $V_{T-}$ values (figure 5), we deduce two values for $\varepsilon_0$ (0.43 and 0.53 eV). These two values can be ascribed to the fact that the **NaphSH** SAM is more disordered and these values should correspond to the LUMO for two phases with different molecular organization co-existing in the SAM. This disorder is also reflected in the C-AFM analysis for which we observe three peaks of conductance negative bias (figure 3c) and the larger distribution of the rectification ratio (see discussion above and figure S7 in SI). Similarly, the IPES spectrum shows two marked "bumps" emerging distinctly from the background in a more pronounced way than for **3TSH** SAM and bare Au (Figure 6 and Figure S10), with LUMO positions extrapolated at about 0.4 and 1.6 eV. While not strictly comparable, the TVS and IPES data are qualitatively in agreement. We deduce only one value of electron affinity (EA) from CV measurements ($EA_{CV}$= -3.75 eV) which may be viewed as an average value compared to the two values measured by IPES (-3.30 and -4.50 eV).

From the Bâldea model of TVS, we can also calculate γ, a voltage division parameter describing the degree of symmetry or asymmetry of the molecular orbitals in the junction (-0.5 ≤ γ ≤ 0.5, γ = 0 being the case of a symmetrical coupling of the molecular orbitals between the two electrodes).[48,54-55] γ is equal to -0.5 if the molecule is fully coupled with the Au substrate and not with the tip, and γ is 0.5 in the opposite situation (see Figure S8 in SI).

$$\gamma = \frac{\text{sign } \varepsilon_0}{2} \frac{V_{T+} + V_{T-}}{\sqrt{V_{T+}^2 + 10|V_{T+}V_{T-}|/3 + V_{T-}^2}} \qquad \text{(Eq. 2)}$$

where sign $\varepsilon_0$ is "+" for LUMO and "-" for HOMO. The values are given in Table 1. Clearly, the value $\gamma = -0.14$ for **3TSH** junctions confirms the asymmetry of this junctions ($R_- = 7.8$ on average for STM I-V measurements) with the terthiophene HOMO more coupled to the Au substrate than the STP tip ($\Gamma_{STMtip} < \Gamma_{Au}$). For the **C$_{12}$SH** and **NaphSH**, $\gamma$ is almost 0. This is expected from the symmetrical I-V behavior of the **C$_{12}$SH** junctions. For the **NaphSH** junctions, it means that the model is not "accurate" enough to describe the weaker rectification observed for the **NaphSH** junctions ($R_+ < 3$ on average, both from C-AFM and STM measurements).

To summarize the electrical properties of SAMs assessed by C-AFM and STS, the chemical nature of the conjugated system, associated to the asymmetrical coupling of the molecular orbitals and the electrodes, involves an opposite rectification. This is explained by the electronic transport through the HOMO or the LUMO levels of the **3TSH** and **NaphSH** which have the redox properties to be respectively oxidized or reduced (see cyclic voltammetry part). Comparing the conductance histograms of the two molecular junctions, those for the **3TSH** junctions are less dispersed (Figure 3). As mentioned above, ~80% of the **3TSH** junctions belong to a main peak, while three distinct peaks are clearly observed for the **NaphSH** junctions. Such a behavior is also confirmed by the transition voltage and rectification ratio distributions - i.e., narrower for **3TSH** than for **NaphSH** - obtained from STM I(V) analyses. This feature may be related to the better organization of the **3TSH** SAMs as observed by high resolution STM images (figure 2) or by CVs with π-π interactions between tertiophene units that are not present in naphthalene derivative. For instance, it was shown that disordered SAMs of alkylthiols on Au lead to junction with higher current than those having a more ordered organization,[56] or that "gauche" defects in the molecule reduce the electron transfer efficiency through the molecules.[57]

We have also observed[8] that a reduced number of peaks (i.e. 2 peaks) for alkylthiol junctions on nanodot (single crystal) electrodes compared to alkylthiol junction on substrate Au (poly-crystalline) electrode (3 peaks) is consistent with the fact that the nanodot size (5-8 nm) is of the same order of magnitude as (or even smaller than) the known average size for well-organized, close-packed, domains in SAM as measured by grazing-angle X-ray diffraction (coherence length of the diffraction peak of about 7 nm for alkyl chain with 18 carbon atoms).[58] It is likely that the SAMs are more disordered in the case of Au-substrate. In all cases, a decrease of the number of peaks is the fingerprint of a better control of the molecular organization in the molecular junctions. While all these observations were done for aklylthiols, it is likely that they also hold in the present case, supporting the conclusion of a better molecular organization for the **3TSH** junctions.

**Conclusion**

The steric hindrance of the conjugated backbone in SAMs on gold was studied through the synthesis and the characterization of **3TSH** and **NaphSH**. The cyclic voltammetry study shows clearly π-π interactions between terthiophene cores in SAMs unlike the naphthalene carbodiimide cores that present no interactions. The density of **3TSH** is twice as large as that of **NaphSH**. I/V characteristics measured by STS on bulk gold substrate or C-AFM on gold nanoplots highlight (i) that the dispersion of the measurements is less important on more organized SAMs as **3TSH** SAMs, (ii) an opposite rectification explained by the chemical nature of the conjugated system and by the asymmetrical coupling of the molecular orbitals and the electrodes, and (iii) that more than 80% of the measurements show a transport through an orbital of the conjugated system despite a worse organization.

ASSOCIATED CONTENT

**Supporting Information**. Synthesis and characterizations of **3TSH** (scheme 1) and **NaphSH** (scheme 2). 3D chemical structures of thiols derivatives modeled by Gaussian software (figure S1). Monolayer characterizations (XPS, Figure S2 and Table S1; Infrared spectroscopy, Figure S3; Cyclic voltammetry, Figure S4, S5 and Table S2). C-AFM measurements on SAMs on gold (Figure S6 and Table S3) and distribution of rectification ratios measured from STM experiments (Figure S7). Schematic representation of transport properties via the HOMO and LUMO levels of organic compounds (Figure S8). Typical TVS curves (Figure S9). Background-subtracted IPES spectra (Figure S10).

AUTHOR INFORMATION

**Corresponding Author**

*Email: lionel.patrone@im2np.fr, stephane.lenfant@iemn.univ-lille1.fr and bruno.jousselme@cea.fr**Notes**

The authors declare no competing financial interests.

ACKNOWLEDGMENT

We are indebted to the French Ministry of Research for the financial support, through the ANR SAGe III-V, Programme Blanc SIMI 10 (Project ANR-11-BS10-012), and support from "Solutions Communicantes Sécurisées (SCS)" international cluster is acknowledged. We also thank the Nanoscience Competence Center of Région Ile-de-France C'Nano IdF (project ''PIGe III-V") for the post-doctoral grant assigned to OS. The "Objectif 2" EEC program (FEDER), the "Conseil General du Var" Council, the PACA Regional Council, Toulon Provence Méditerranée and ISEN-Toulon are acknowledged for equipment funding. The authors thank M. Abel, M. Koudia, and J.M. Themlin for UPS/IPES experiments and discussions, N. Clément for the fabrication of the gold nanodot arrays and help with the statistical data analysis.

# Influence of Molecular Organization on the Electrical Characteristics of π-conjugated Self-assembled Monolayers.

Xavier Lefèvre,[1] Fabrice Moggia,[1] Olivier Segut,[1] Yu-Pu Lin,[2] Younal Ksari,[2] Grégory Delafosse,[2,3] Kacem Smaali,[4] David Guérin,[4] Vincent Derycke,[1] Dominique Vuillaume,[4] Stéphane Lenfant,[3*] Lionel Patrone,[2,3*] Bruno Jousselme[1*]

AUTHOR ADDRESS

[1] CEA Saclay, IRAMIS, NIMBE/UMR 3685, Laboratory of Innovation in Surface Chemistry and Nanosciences (LICSEN), Gif-sur-Yvette Cedex F-91191, France

[2] Institut Matériaux Microélectronique Nanosciences de Provence, Aix-Marseille Université, CNRS, Université de Toulon, IM2NP UMR 7334, Domaine Universitaire de St Jérôme, Service 151, 13397, Marseille Cedex 20, France

[3] Institut Supérieur de l'Electronique et du Numérique (ISEN-Toulon), CNRS, IM2NP UMR 7334, Maison des Technologies, Place Georges Pompidou, F-83000, Toulon, France

[4] Institute for Electronic Microelectronics and Nanotechnology (IEMN), CNRS, Univ. Lille, Av. Poincaré, F-59652 cedex, Villeneuve d'Ascq, France.

**Chemical synthesis of 3TSH and NaphSH.**

Synthesis of 5-hexylsulfanyl-5''-(6-mercaptohexylsulfanyl)-2,2′:5′,2″-terthiophene (**3TSH**) described previously for bithiophene core[1] is outlined in scheme S1. Compound **1** was prepared in a one-pot reaction on terthiophene in anhydrous THF with two equivalent of *n*-butyllithium at low temperature and insertion of elemental sulfur to generate the corresponding dithiolates which were alkylated with 3-bromopropionitrile. The mono deprotection of one thiolate group of **1** with action of 1 equivalent of CsOH and treatment with bromohexane gave **2**. At this step, the **3THex** coming from the complete deprotection was also isolated during column chromatography; it will be used as reference compound in the electrochemistry study. The same procedure is followed to remove the second cyanoethyl group and the treatment with *S*-6-bromohexyl ethanethioate[2] gave **3**. The final deprotection of thiol group is realized by reduction of the thioester with diisobutylaluminum hydride in $CH_2Cl_2$ to afford the target **3TSH**.

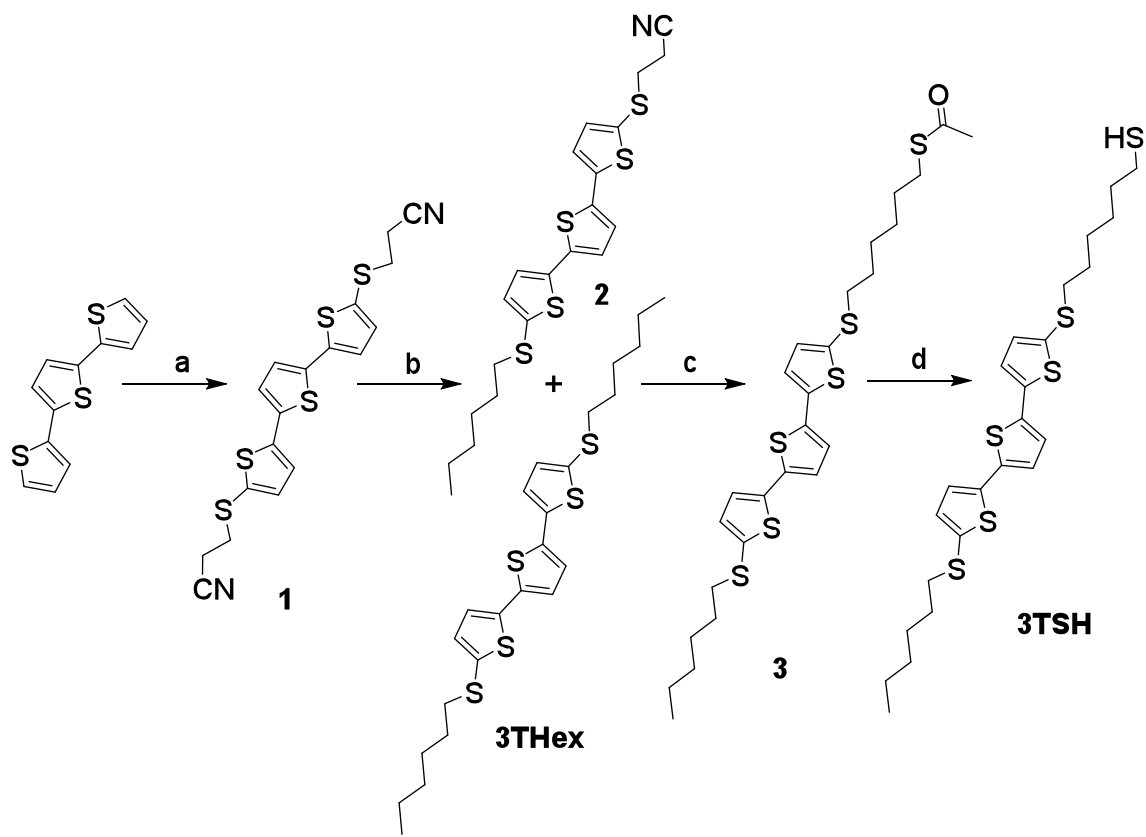

**Scheme S1**. Synthesis of **3TSH**. a) 1. *n*-BuLi/THF, -78°C, 2. $S_8$, 3. $NC(CH_2)_2Br$, $N_2$ (42%). b) 1. CsOH/MeOH, DMF, 2. 1-Bromohexane, RT, $N_2$ (20% for **3THex** and 37% for **2**). c) 1. CsOH/MeOH, DMF, 2. *S*-6-bromohexyl ethanethioate, RT, $N_2$ (71%). d) 1. Diisobutylaluminum hydride/$CH_2Cl_2$, 0°C, $N_2$ 2. HCl (81%).

The target *N*-(6-mercaptohexyl)-*N'*-hexylylnaphthalene-1,8:4,5-tetracarboxydiimide (**NaphSH**) was synthesized according to the procedure depicted in Scheme S2. Compound **4** was prepared as previously described in literature from dibromohexane. The introduction of thioester group was done by action of potassium thioacetate on **4** in dry THF. The deprotection of phtalimide and thioacetate group was done in a one-pot reaction by using four equivalents of hydrazine mono-hydrate in absolute ethanol to give the desired aminothiol. The coupling of **6** with N-Hexylnaphthalene-1,8-dicarboxyanhydride-4,5-dicarboximide in DMF at 120°C during one night gives N-(6-mercaptohexyl)-N'-hexylylnaphthalene-1,8:4,5-tetracarboxydiimide (**NaphSH**).

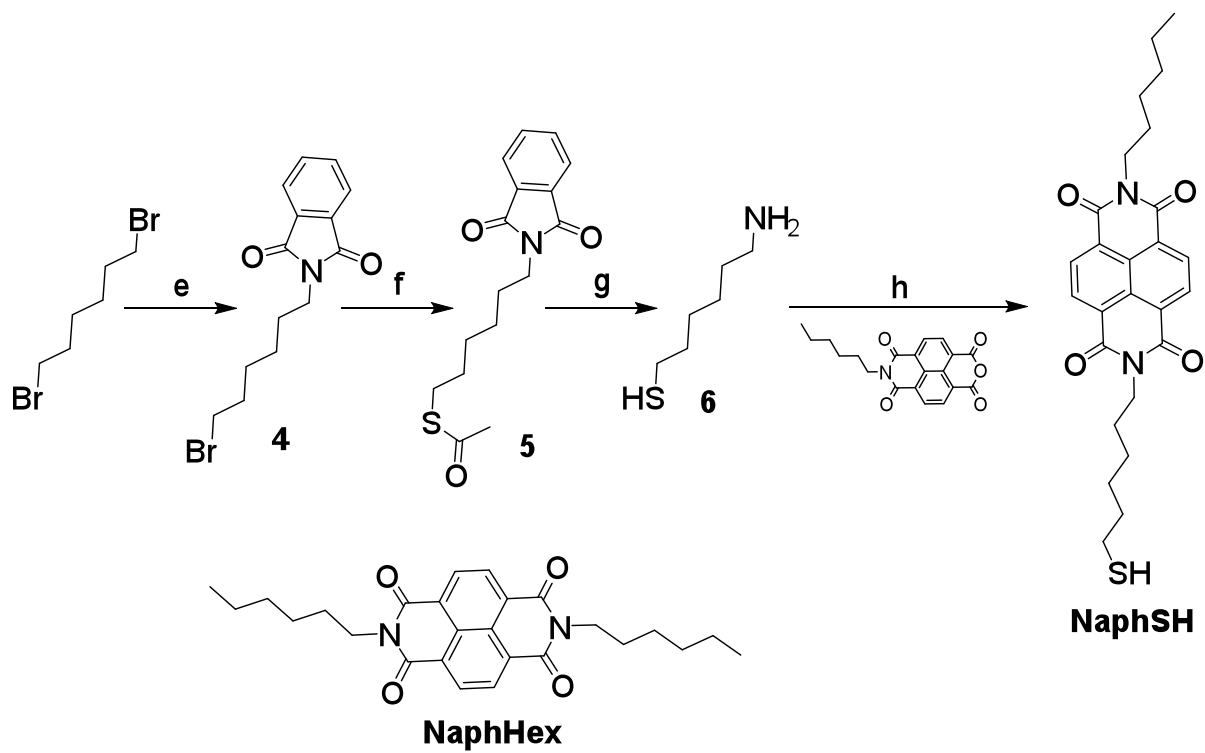

**Scheme S2.** Synthesis of **NaphSH**. e) Potassium phtalimide, Acetone, reflux (47%). f) Potassium thioacetate, THF, reflux (57%). g) Hydrazine, EtOH, reflux (32%). h) DMF, 120°C, $N_2$ (41%). Chemical structure of **NaphHex**.

**Experimental details:**

All reagents and chemicals were purchased from Aldrich and were used as received. All reactions were performed under inert atmosphere using anhydrous solvents which were dried and distilled before being used. S-6-bromohexyl ethanethioate,[2] N,N'-Dihexylnaphthalene-1,8-dicarboxyanhydride-4,5-dicarboximide[3] and N-Hexylnaphthalene-1,8-dicarboxyanhydride-4,5-dicarboximide[4] were prepared according to procedures described previously. Melting points are uncorrected and were obtained from a Stuart SMP3 apparatus. NMR spectra were recorded with a Bruker AVANCE DRX 400 (1H, 400 MHz and $^{13}$C, 100 MHz). Chemical shifts δ are expressed in ppm relative to tetramethylsilane (TMS). UV-Vis spectra were recorded with a Perkin Elmer Lambda 650 spectrometer. The mass spectrometry analysis was operated with a Thermo Finnigan Automass in the electron impact mode at 50 or 70 eV ion source energy equipped with a quadripolar analyser. Elemental analyses were done at the Microanalyses Laboratory at ICSN/CNRS, Gif/Yvette, France.

**5,5''-Bis(2-cyanoethylsulfanyl)-2,2':5',2''-terthiophene 1.** To a solution of 2,2':5',2''-terthiophene (1.00 g, 4 mmol) in anhydrous THF (25 mL) under $N_2$ at –78°C was added dropwise a solution of *n*-BuLi 2.5 M in hexane (3.4 mL, 2.1 eq.) over a period of 5 min. After 30 min of stirring at –78°C, elemental sulfur (270 mg, 2.1 eq.) was added in one portion to the yellow precipitate obtained. After 30 min of additional stirring at –78°C, the red reaction mixture was allowed to warm to room temperature and stirred for 30 min. Then the solution was cooled down to 0°C and 3-bromopropionitrile (1.35 g, 2.5 equiv) was added dropwise. The reaction mixture was stirred for 1 h at room temperature to give a yellow solution. After addition of a 0.5 M aqueous solution of HCl (40 mL), the mixture was extracted with $CH_2Cl_2$. The organic phases were dried over $Na_2SO_4$ and evaporated in vacuo. The chromatography on silica gel (eluent: $CH_2Cl_2$/hexane 1:1 then $CH_2Cl_2$) gave **1** as a yellow solid (700 mg; 42% yield). M.p. = 167 - 168 °C, $^1$H NMR (CDCl$_3$) δ: 7.14 (d, 2H, $^3J$=3.7), 7.08 (s, 1H), 7.07 (d, 2H, $^3J$= 3.7), 3.01 (t, 4H, $^3J$=7.1), 2.66 (t, 4H, $^3J$=7.1). FTIR ν/cm$^{-1}$: 3063, 2249, 1494, 1454, 1425, 1407, 1318, 1280, 1219, 1206, 1193, 1163, 1074, 1054, 983, 956, 904, 849, 787. UV/Vis (CHCl$_3$) : λ = 378 nm. EI MS (50 eV) m/z (I%): 418 (M$^{+\cdot}$, 57), 364 (100), 310 (25), 233 (15), 162 (10). Elem. Anal.: calculated for $C_{18}H_{14}N_2S_5$ (C 51.64%, H 3.32%, S 38.46%), found (C 51.85%, H 3.27%, S 38.01%).

**5-(2-cyanoethylsulfanyl)-5''-hexyl-2,2':5',2''-terthiophene 2** and **5,5''-dihexyl-2,2':5',2''-terthiophene (3THex)**. To a solution of **1** (625 mg, 1.5 mmol) in degassed DMF (50 mL) was added dropwise a solution of CsOH·H$_2$O (0.25 g, 1 equiv.) in degassed MeOH (5 mL) under N$_2$ atmosphere. After additional stirring of 1 H, 1-Bromohexane was added to the red solution. The solution then returns to this initial yellow color after 1 H of stirring. Water (100 ml) was added to the mixture and the compound was extracted with CH$_2$Cl$_2$ (200 mL) and then washed with water. After drying (Na$_2$SO$_4$), the solution was concentrated in vacuo giving a residue which was chromatographed on silica gel (CH$_2$Cl$_2$/Hexane (1/1) then CH$_2$Cl$_2$/Hexane (8/2)) to give first **3THex** (142 mg, 20%) and secondly **2** (250 mg, 37%) as yellow solids. **3THex** : M.p. = 73 - 74 °C. $^1$H NMR (CDCl$_3$) δ: 7.02 (s, 2H), 7.01 (d, 2H, $^3J$=3.7), 7.00 (d, 2H, $^3J$=3.7), 2.82 (t, 4H, $^3J$=7.3), 1.64 (qu, 4H, $^3J$=7.2), 1.41 (qu, 4H, $^3J$=7.2), 1.35-1.23 (m, 8H), 0.89 (t, 6H, $^3J$=7.0). $^{13}$C NMR (CDCl$_3$) δ: 139.9, 136.0, 134.5, 133.9, 124.3, 123.7, 38.9, 31.3, 29.4, 28.1, 22.5, 14.0. FTIR ν/cm$^{-1}$: 3061, 2953, 2923, 2870, 2855, 1495, 1459, 1426, 1382, 1314, 1254, 1217, 1200, 1078, 1050, 986, 852, 783, 729, 721. UV/Vis (CHCl$_3$) : λ = 386 nm. **2** : M.p. = 100 - 102 °C. $^1$H NMR (CDCl$_3$) δ: 7.13 (d, 1H, $^3J$=3.7), 7.07-7.01 (m, 4H), 7.00 (d, 1H, $^3J$=3.7), 3.00 (t, 2H, $^3J$=7.1), 2.82 (t, 2H, $^3J$=7.5), 2.66 (t, 2H, $^3J$=7.1), 1.64 (qu, 2H, $^3J$=7.4), 1.41 (m, 2H), 1.30 (m, 4H), 0.89 (t, 3H, $^3J$=7.0). FTIR ν/cm$^{-1}$: 3060, 2953, 2923, 2870, 2854, 2249, 1493, 1454, 1425, 1410, 1380, 1318, 1280, 1200, 1075, 985, 850, 788. UV/Vis (CHCl$_3$) : λ = 384 nm. EI MS (50 eV) m/z (I%): 449 (M$^+$, 66), 395 (73), 364 (39), 310 (20), 43 (100). Elem. Anal.: calculated for C$_{21}$H$_{23}$NS$_5$ (C 56.08%, H 5.15%, S 35.65%), found (C 56.26%, H 5.18%, S 35.64%).

**5-hexyl-5''-(8-Oxo-7-thianonylsulfanyl)-2,2':5',2''-terthiophene 3.** To a solution of **2** (500 mg, 1.11 mmol) in degassed DMF (15 mL) was added dropwise a solution of CsOH·H$_2$O (217 mg, 1.3 equiv.) in degassed MeOH (5 mL) under N$_2$ atmosphere. After additional stirring of 1 H, thioacetic acid *S*-(6-bromohexyl) ester[2] (0.8 g, 3.33 mmol, 3 equiv.) diluted in degassed DMF (10 mL) was added to the red solution. The solution then returns to the yellow initial color, after 2 H of stirring DMF was removed under vacuum to give a crude product which was chromatographed on silica gel (CH$_2$Cl$_2$/Hexane (7/3) to give a yellow powder (440 mg, 71%). M.p. = 70 - 72 °C, $^1$H NMR (CDCl$_3$) δ : 7.03 (s, 2H), 7.01 (d, 2H, $^3J$=3.7), 7.00 (d, 2H, $^3J$=3.7), 2.86 (t, 2H, $^3J$=7.3), 2.82 (t, 2H, $^3J$=7.3), 2.82 (t, 2H, $^3J$=7.3), 2.32 (s, 3H), 1.66-1.59 (m, 6H), 1.43-1.40 (m, 6H), 1.30-1.28 (m, 4H), 0.88 (t, 3H, $^3J$=7.1). FTIR ν/cm$^{-1}$: 3061, 2951, 2923, 2870, 2854, 1700, 1495, 1461, 1425, 1380, 1352, 1313, 1265, 1146, 1118, 985, 955, 851, 786. UV/Vis (CH$_2$Cl$_2$) : λ = 384 nm. EI MS (70 eV) m/z (I%): 554 (M$^{+\cdot}$, 100), 469 (25), 395 (66), 311 (20), 117 (17). Elem Anal: calculated for C$_{26}$H$_{34}$OS$_6$ (C 56.27%, H 6.18%, S 34.67%) found (C 56.49%, H 6.19%, S 34.65%).

**5-Sulfanylhexyl-5''-(6-mercaptohexylsulfanyl)-2,2':5',2''-terthiophene (3TSH).** To a solution of **3** (440 mg, 0.80 mmol) in anhydrous CH$_2$Cl$_2$ (15 mL) was added dropwise a solution of diisobutylaluminum hydride (228 mg, 2 equiv.) in CH$_2$Cl$_2$ (5 mL) under N$_2$ atmosphere. After 2 hours, HCl solution (1M, 20 mL) was added to the mixture and the organic layer was isolated, washed with water, dried with Na$_2$SO$_4$ and evaporated. The chromatography on silica gel (eluent: CH$_2$Cl$_2$/hexane 3:7) gave **3TSH** as a yellow solid (330 mg; 81% yield). M.p. = 195 - 197 °C, $^1$H NMR (CDCl$_3$) δ : 7.03–6.98 (m, 6H), 2.81 (t, 4H, $^3J$=7.3), 2.50 (m, 2H), 1.71-1.59 (m, 6H), 1.43-1.36 (m, 6H), 1.32-1.23 (m, 4H), 0.88 (t, 3H, $^3J$=7.0). $^{13}$C NMR (CDCl$_3$) δ : 134.1 (2C), 133.9 (2C), 124.4 (2C), 124.3 (2C), 123.8 (4C), 38.9, 38.8, 33.8, 31.3, 29.4, 29.2, 28.1, 27.8 (2C), 24.5, 22.5, 14.0. FTIR ν/cm$^{-1}$: 3060, 2951, 2923, 2871, 2854, 1495, 1461, 1425, 1380, 1313, 1269, 1217, 1200, 1190, 1078, 1050, 985, 851, 787. UV/Vis (CH$_2$Cl$_2$) : λ = 385 nm (log ε =

4.30). EI MS (50 eV) m/z (I%): 512 (M$^{+\cdot}$, 30), 395 (34), 311 (15), 44 (100). Elem Anal: calculated for C$_{24}$H$_{32}$S$_6$ (C 56.20%, H 6.29%, S 37.51%) found (C 56.48%, H 6.21%, S 36.88%).

**_N_-(6-bromohexyl)phtalimide (4)**. Potassium phthalimide salt (1.9 g, 0.5 equiv.) was added in 4 parts during 6 h to a solution of 1,6-dibromohexane (3.15 mL, 20.5 mmol) dissolved in boiling acetone (64 mL). After an additional 24 h reaction time at reflux, the solution was cooled to room temperature. The white solid obtained was removed by filtration and the filtrate was concentrated under vacuum to give a crude product which was chromatographed on silica gel (Hexane/Et$_2$O (6/4) to give **4** as a solid (2.95 g, 47%). M.p. : 55 - 57°C. NMR $^1$H (CDCl$_3$) δ : 7.80 (m, 2H), 7.70 (m, 2H), 3.61 (t, 2H, $^3J$=7.1), 3.36 (t, 2H, $^3J$=6.8), 1.87–1.79 (m, 2H), 1.71–1.63 (m, 2H), 1.50–1.42 (m, 2H), 1.37–1.30 (m, 2H). FTIR ν/cm$^{-1}$: 2960, 2931, 2855, 1769, 1707, 1614, 1433, 1394, 1358, 1330, 1233, 1213, 1049, 973, 890, 711.

**_N_-(8-Oxo-7-thianonyl)phtalimide (5).** Potassium thioacetate (736 mg, 1 equiv.) was added in one portion to a solution of **4** (2g, 6.45 mmol) diluted in freshly distilled THF (20 mL). Then the mixture was heated under reflux over 24 hours. After cooling at room temperature, the white precipitate was removed by filtration and the solvent evaporated under reduced pressure. The crude product was purified by chromatography on silica gel (Hexane/Et$_2$O (6/4)) to give a solid (1.12 g, 57%). M.p. : 60 - 61°C. NMR $^1$H (CDCl$_3$) δ : 7.83 (m, 2H), 7.70 (m, 2H), 3.67 (t, 2H, $^3J$=7.3), 2.84 (t, 2H, $^3J$=7.3), 2.30 (s, 3H), 1.71–1.64 (m, 2H), 1.60–1.52 (m, 2H), 1.44–1.30 (m, 4H). FTIR ν/cm$^{-1}$: 2959, 2938, 2855, 1765, 1693, 1680, 1463, 1434, 1397, 1361, 1337, 1117, 1044, 1012, 993, 957, 904, 716.

**6-aminohexylthiol (6).** To a solution of **5** (1g, 3.28 mmol) in ethanol (30 mL) was added hydrazine hydrate (0.41 mL, 4 eq.), then the solution was heated under reflux overnight. After cooling to room temperature the solvent was removed under reduced pressure. Methylene chloride (20 mL) was added to the crude product and the organic layer was washed 3 times with 20 mL of water, dried over magnesium sulfate, and concentrated under reduced pressure. The colorless oil of the aminohexylthiol **6** was directly used in the following step without further purification (141 mg, 32%). $^1$H NMR (CDCl$_3$) : δ = 2.64-2.70 (m, 4H), 1.72–1.63 (m, 2H), 1.56 (s, 2H), 1.48 - 1.29 (m, 7H).

**N-(6-mercaptohexyl)-N'-hexylnaphthalene-1,8:4,5-tetracarboxydiimide (NaphSH).** A solution of **6** (70 mg, 0.525 mmol) and *N*-Hexylnaphthalene-1,8-dicarboxyanhydride-4,5-dicarboximide[4] (185 mg, 0.525 mmol) in DMF (10 mL) was heated at 120°C under N$_2$ overnight. After cooling, a pinkish solid was formed and removed by filtration. This solid was washed several times with pentane to give 100 mg (41%) of **NaphSH**. M.p. = 237 - 238 °C, $^1$H NMR (CDCl$_3$) δ : 8.74 (s, 4H), 4.19 (t, 4H, $^3J$=7.6), 2.50 (m, 2H), 1.80–1.67 (m, 6H), 1.50-1.40 (m, 6H), 1.40-1.31 (m, 4H), 0.90 (t, 3H, $^3J$=7.0). $^{13}$C NMR (CDCl$_3$) δ : 162.8 (4C), 130.9 (4C), 126.6 (4C), 126.5 (2C), 41.0, 40.8, 38.9, 31.5, 29.0, 28.1, 28.0, 27.9, 26.7, 26.6, 22.5, 14.0. FTIR ν/cm$^{-1}$: 3061, 2929, 2853, 1704, 1655, 1581, 1454, 1375, 1333, 1246, 1185, 1154, 1077, 971, 890, 769, 726. UV/Vis (CH$_2$Cl$_2$) : λ = 380 nm (log ε = 4.39), 360 nm (log ε = 4.32), 340 nm (log ε = 4.07). EI MS (50 eV) m/z (I%): 466 (M$^{+\cdot}$, 7), 351 (8), 281 (3), 249 (5), 115 (25), 55 (40), 44 (100). Elem Anal: calculated for C$_{26}$H$_{30}$N$_2$O$_4$S (C 66.93%, H 6.48%, N 6.00%, S 6.87%) found (C 66.95%, H 6.18%, N 6.01%, S 6.91%).

**Modelisation**

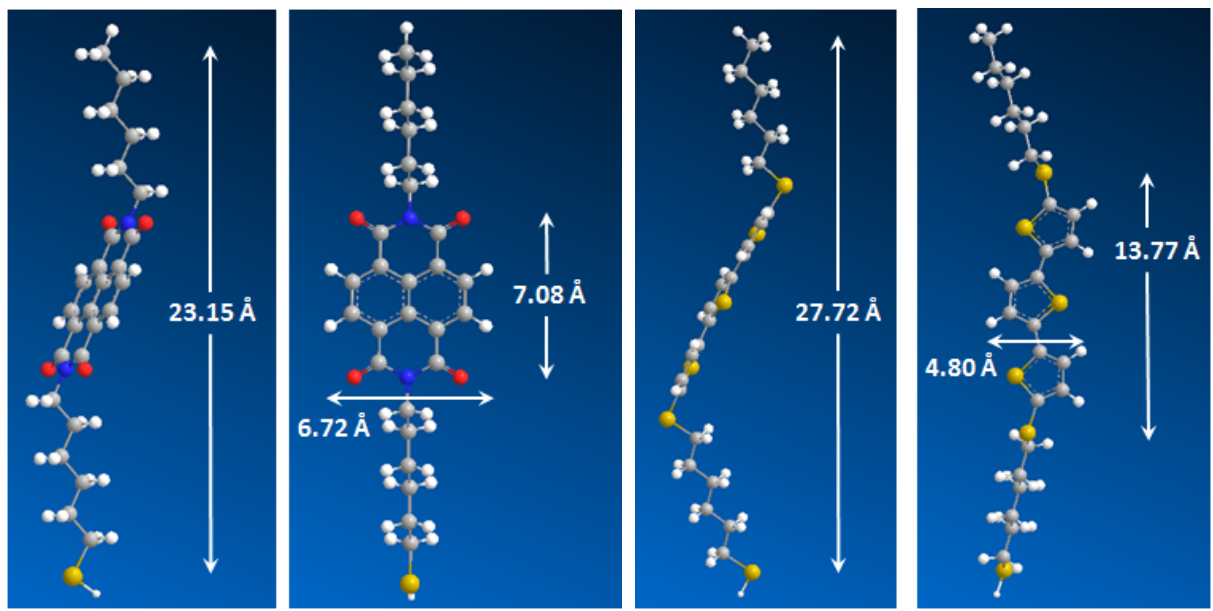

**Figure S1.** Geometry optimized of **NaphSH** (left) and **3TSH** (right) by the functional density method Becke3lyp using Gaussian 98.[5] Bases for calculation are 6-31G*.

**XPS analysis**

X-ray photoelectron spectroscopy (XPS) was performed with a Kratos Analitical Axis Ultra DLD, using an Al Ka source monochromatized at 1486.6 eV. A hemispheric analyzer working at pass energy of 50 eV for the global spectrum was used, and 20 eV when focusing on the sole core levels.

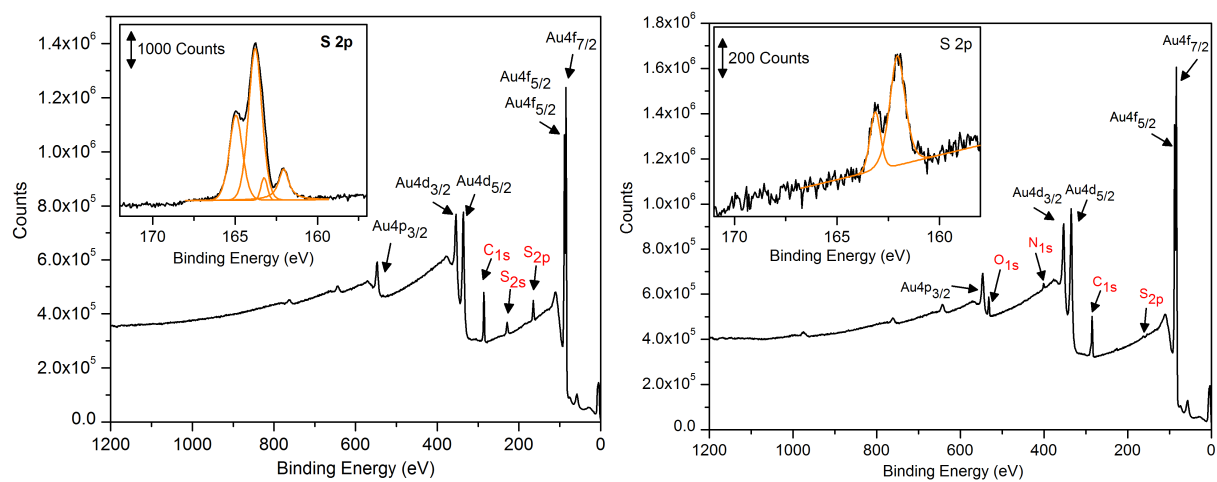

**Figure S2.** XPS spectra of **3TSH** (left) and **NaphSH** (right) on gold substrates and the insets show the S 2p core level.

**Table S1**. Ratio analysis of the peaks in XPS spectra in monolayers of **3TSH** and **NaphSH** on gold.

| **3TSH** | n | % found | % calcd | **NaphSH** | n | % found | % calcd |
|---|---|---|---|---|---|---|---|
| C | 24 | 79.3 | 80.0 | C | 26 | 78.8 | 78.8 |
| S | 6 | 20.7 | 20.0 | N | 2 | 5.5 | 6.0 |
|   |   |   |   | O | 4 | 12.4 | 12.1 |
|   |   |   |   | S | 1 | 3.3 | 3.1 |

n corresponds to the number of atoms in the chemical structure.

**IR analysis**

Infrared spectroscopy (IR) was carried out with a Bruker Vertex 70 spectrometer (resolution 2 cm$^{-1}$, spectra were collected with 256 scans, MCT detector) equipped with a Pike Miracle plate for ATR.

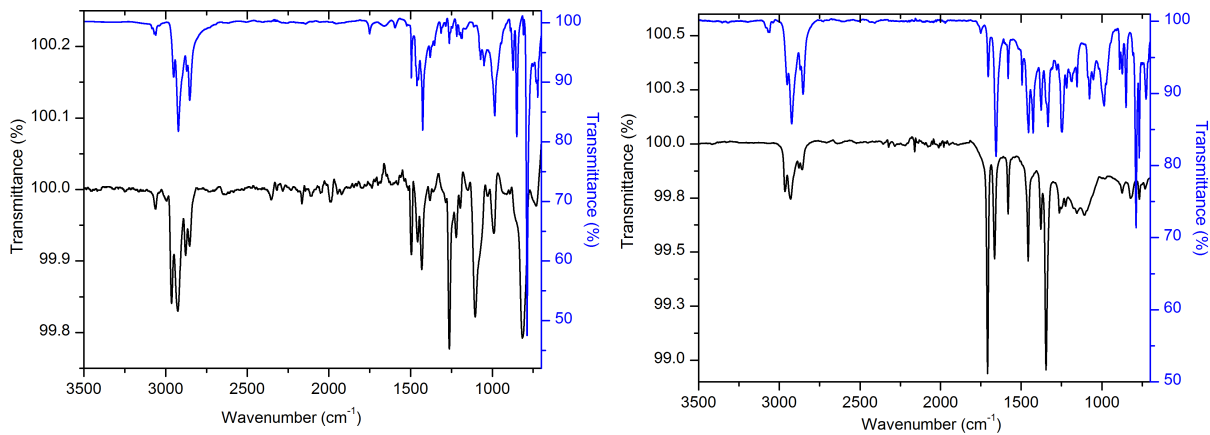

**Figure S3**. IR spectra of **3TSH** (left) and **NaphSH** (right) in carbon tetrachloride solution (blue curve) and chemisorbed on gold surface (black curve).

**Electrochemical study.**

Cyclic voltammetry was performed in acetonitrile solutions (HPLC grade). Tetrabutylammonium hexafluorophosphate (0.1 M as supporting electrolyte) from Sigma-Aldrich was used without purification. Experiments in a glovebox were done in Plate material evaluating cell from Biologic with a gold working and a platinum wire as counter electrode. An Ag/Ag$^+$ (10 mM) electrode checked against the ferrocene/ferricinium couple (Fc/Fc$^+$) before and after each experiment was used as reference. The reference electrode was equipped with a non aqueous double bridge in order to avoid possible interference with metal cations. Electrochemical experiments were carried out with an EG&G potentiostat, model 273A.

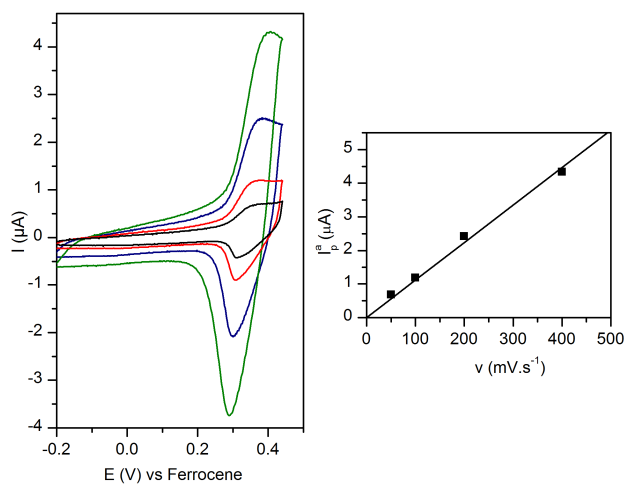

**Figure S4**. left: Cyclic voltammograms of **3TSH** SAM-Au electrode recorded at different scan rates (50, 100, 200 and 400 mV.s–1); right: trace of Ipa as a function of the scan rate of **3TSH** SAM-Au electrode. All measurements have been made in $CH_2Cl_2$-$TBAPF_6$ 0.1 mol.$L^{-1}$. Potentials are quoted versus Ferrocene.

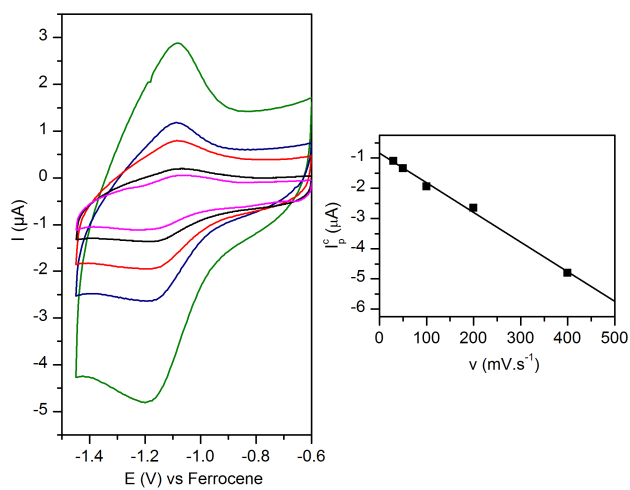

**Figure S5**. left: Cyclic voltammograms of **NaphSH** SAM-Au electrode recorded at different scan rates (30, 50, 100, 200 and 400 mV.s–1); right: trace of Ipc as a function of the scan rate of **NaphSH** SAM-Au electrode. All measurements have been made in $CH_2Cl_2$-$TBAPF_6$ 0.1 mol.$L^{-1}$. Potentials are quoted versus Ferrocene.

**Table S2**. Electrochemical Data (millivolts) of reference compounds in solution (with gold working electrode) and thiols derivatives chemisorbed on gold (0.1 M $Bu_4NPF_6$ / $CH_2Cl_2$; potentials vs Ferrocene).

|  | *E°' red(2)* | *E°' red(1)* | *E°'ox(1)* | *E°'ox(2)* |
|---|---|---|---|---|
| **3TSHex** in solution |  |  | 435 | 560 |
| **3TSH** chemisorbed |  |  | 340 |  |
| **NaphtSHex** in solution | -1550 | -1130 | | |
| **NaphtSH** chemisorbed | | -1125 | | |

**C-AFM measurements.**

Nanodot fabrication: The fabrication and detailed characterization of these nanodot arrays have been reported elsewhere.[6] For e-beam lithography, an EBPG 5000 Plus from Vistec Lithography was used. The (100) Si substrate (resistivity = $10^{-3}$ Ω.cm) was cleaned with UV-ozone and native oxide etched before resist deposition. The e-beam lithography has been optimized by using a 45 nm-thick diluted (3:5 with anisole) PMMA (950 K). For the writing, an acceleration voltage of 100 keV was used, which reduces proximity effects around the dots, compared to lower voltages. The beam current to expose the nanodots was 1 nA. The conventional resist development / e-beam Au evaporation (8 nm) / lift-off processes were used. Immediately before evaporation, native oxide is removed with dilute HF solution to allow good electrical contact with the substrate. Single crystal Au nanodots were obtained after thermal annealing at 260°C during 2 h under $N_2$ atmosphere. At the end of the process, these nanodots were covered with a thin layer of $SiO_2$ that was removed by HF at 1% for 1 mn prior to SAM deposition. Spacing between Au nanodots was set to 100 nm.

C-AFM measurements: Conducting atomic force microscopy (C-AFM) was performed under a flux of $N_2$ gas (Dimension 3100, Veeco), using a PtIr coated tip (same tip for all C-AFM measurements). Tip curvature radius was about 30-40 nm, and the force constant was in the range 0.17-0.2 N/m. The C-AFM measurements were taken at loading forces of 1 nN. Images were acquired with a sweep frequency of 0.5 Hz and the voltage applied on the substrate.

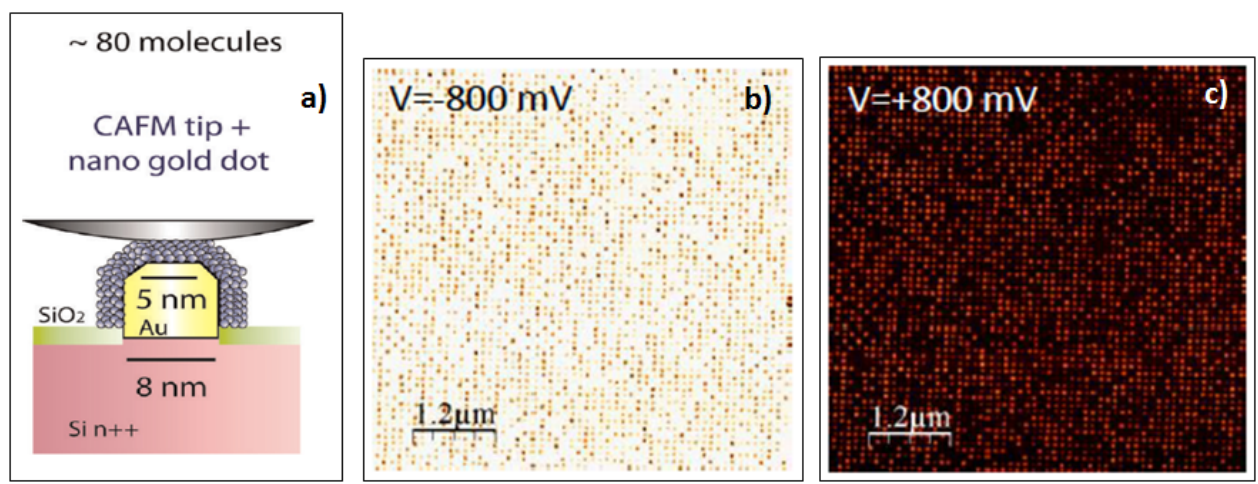

**Figure S6**. (a) Schematic of the nanodot molecular junctions (b) Current image at -0.8V (black spot = negative current and (c) current image at +0.8V (white spot = positive current) of Au nanodot/**NaphSH**/C-AFM tip molecular junctions.

**Table S3**. Parameters of the fitted log-normal distributions for **NaphSH** and **3TSH** at -0.8 V and 0.8V.

| NaphSH | | log I | I (A) | log σ | σ |
|---|---|---|---|---|---|
| @ -0.8V | LC | -9.14 | $7.2 \times 10^{-10}$ | 0.12 | 1.3 |
| | MC | -8.86 | $1.4 \times 10^{-9}$ | 0.22 | 1.6 |
| | HC | -8.62 | $2.4 \times 10^{-9}$ | 0.08 | 1.2 |
| @ +0.8V | | -8.64 | $2.3 \times 10^{-9}$ | 0.25 | 1.8 |
| 3TSH | | | | | |
| @ -0.8V | | -8.47 | $3.4 \times 10^{-9}$ | 0.04 | 1.1 |
| @ +0.8V | | -8.75 | $1.8 \times 10^{-9}$ | 0.04 | 1.1 |
| $C_{12}SH$ | | | | | |
| @ -0.4V | LC | -8.11 | $7.8 \times 10^{-9}$ | 0.28 | 1.9 |
| | HC | -7.85 | $1.4 \times 10^{-8}$ | 0.07 | 1.2 |
| @ +0.4V | LC | -8.14 | $7.2 \times 10^{-9}$ | 0.19 | 1.5 |
| | HC | -7.91 | $1.2 \times 10^{-8}$ | 0.08 | 1.2 |

**Distribution of rectification ratios measured from STM experiments.**

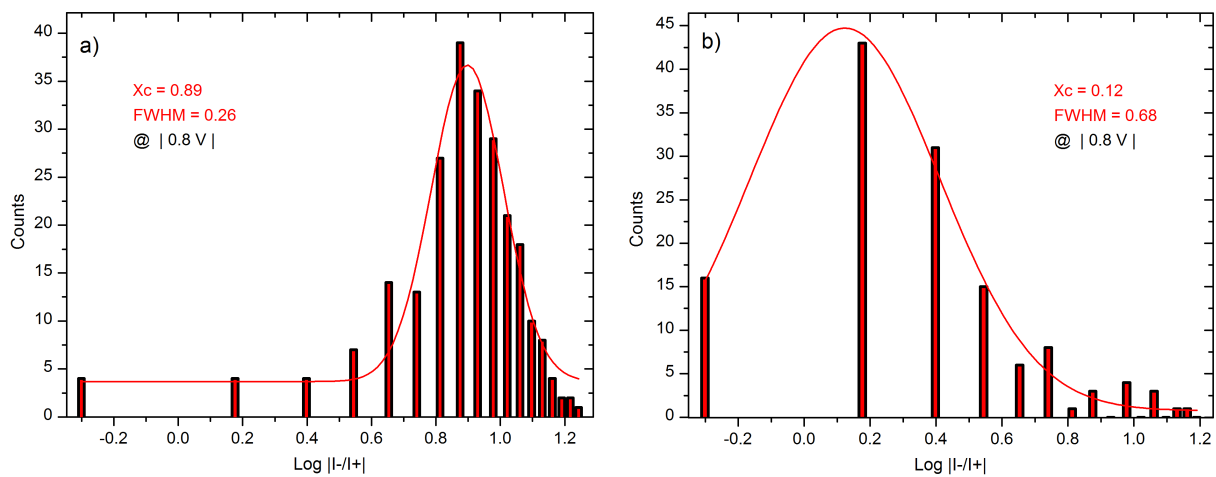

**Figure S7**. Log-normal distributions of rectification ratios R = $|I_-/I_+|$ at ± 0.8 V for **3TSH** (a) and R = $|I_+/I_-|$ at ± 0.8 V for **NaphSH** (b). Xc stands for the maximum position of the Gaussian fit, and full width at half maximum (FWHM) is reported.

**Schematic representation of the electronic transport.**

- Through the HOMO level of the terthiophene derivative.

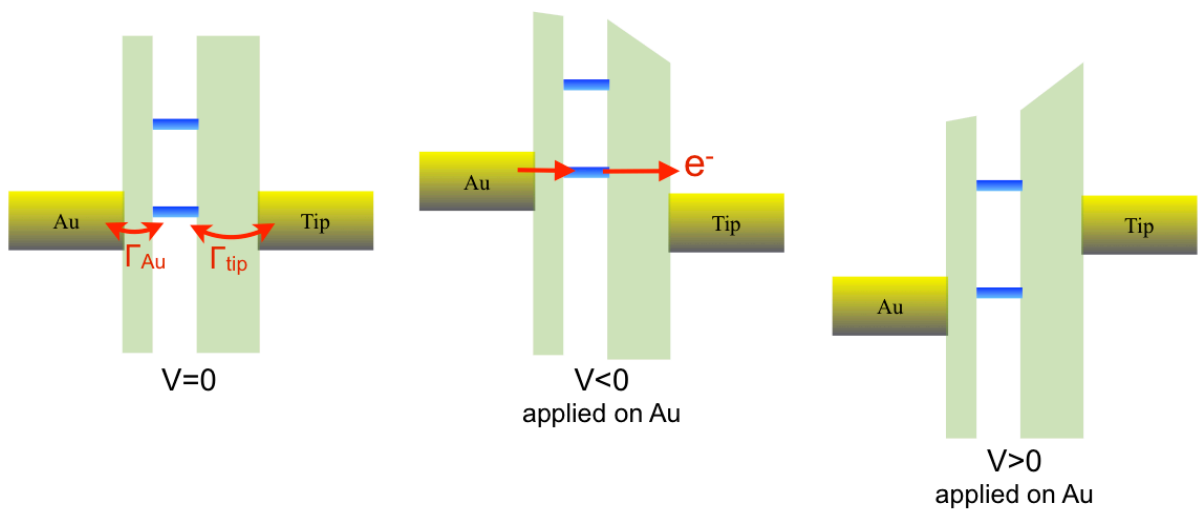

-Through the LUMO level of the tetracarboxydiimide derivative.

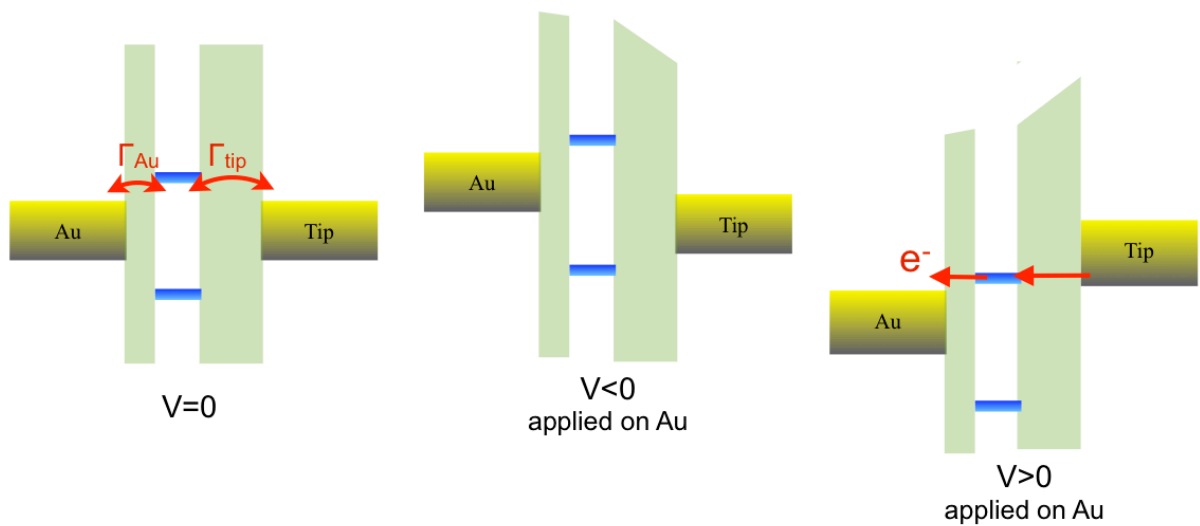

**Figure S8**. Schematic representation of transport properties via the HOMO and LUMO levels of organic compounds in an asymmetrically coupled geometry ($\Gamma_{Au} > \Gamma_{tip}$).

**Typical TVS curves.**

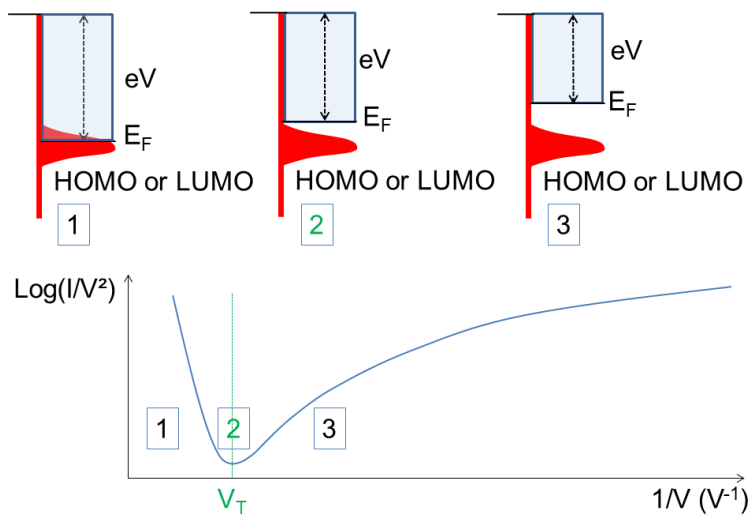

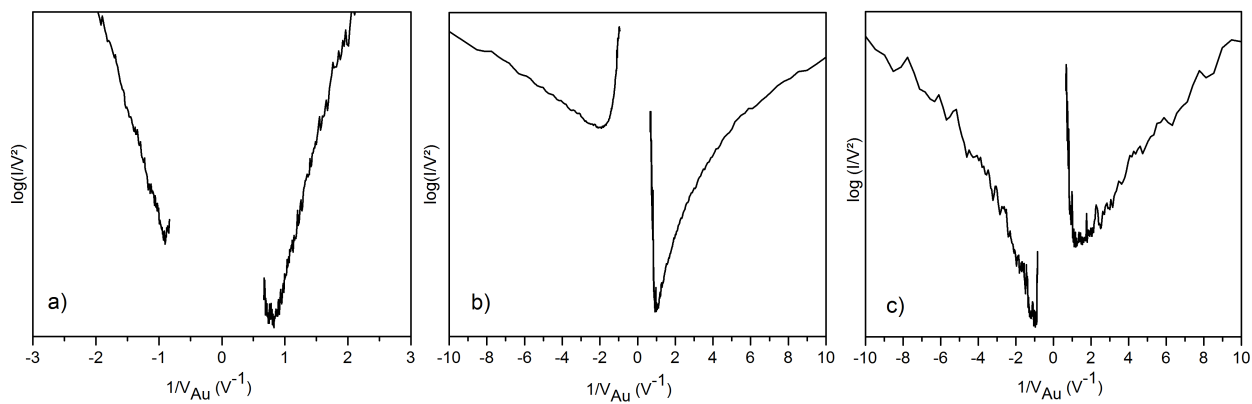

**Figure S9**. Top: principle of the transition voltage determination; the transition voltage at the minimum of the Fowler-Nordheim plot corresponds to the alignment of the Fermi level with the onset of the molecular band tail. Bottom: Typical TVS curves for (a) **C$_{12}$SH** junction, (b) **3TSH** junction and (c) **NaphSH** junction.

**IPES spectra corrected with the background signal**

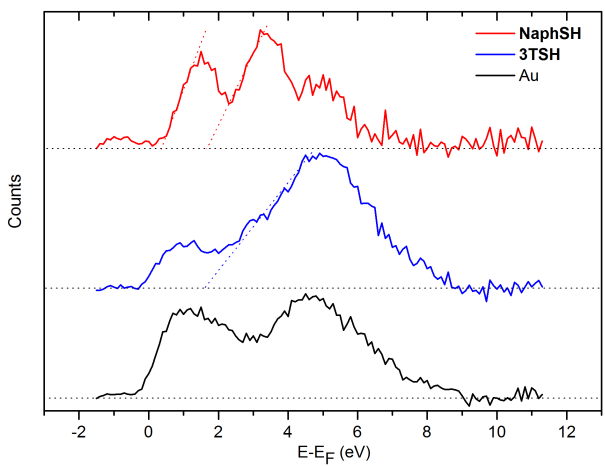

**Figure S10.** IPES spectra with subtracted background signal. Spectra were shifted along the Y axis for the sake of clarity. The onset of the LUMO band tail was determined by the intersection between the tangent to the band (dotted line) and the horizontal baseline.